\documentclass[journal]{IEEEtran}
\usepackage[latin9]{inputenc}
\usepackage{color}
\usepackage{amsmath}
\usepackage{amsthm}
\usepackage{graphicx}
\PassOptionsToPackage{normalem}{ulem}
\usepackage{ulem}
\usepackage[unicode=true]
 {hyperref}

\makeatletter
\theoremstyle{plain}
\newtheorem{thm}{\protect\theoremname}
\theoremstyle{remark}
\newtheorem{rem}[thm]{\protect\remarkname}

\usepackage{xcolor}

\ifCLASSINFOpdf

\else

\fi

\makeatother

\providecommand{\remarkname}{Remark}
\providecommand{\theoremname}{Theorem}

\begin{document}
\title{XR-RF Imaging Enabled by Software-Defined Metasurfaces and Machine
Learning: Foundational Vision, Technologies and Challenges}
\author{C. Liaskos, A. Tsioliaridou, K. Georgopoulos, G. Morianos, S. Ioannidis,
I. Salem, D. Manessis, S. Schmid\\
D. Tyrovolas, S. A. Tegos, P.-V. Mekikis, P. D. Diamantoulakis, A.
Pitilakis, N. Kantartzis, G. K. Karagiannidis\\
A. Tasolamprou, O. Tsilipakos, M. Kafesaki, I.F. Akyildiz, A. Pitsillides,
M. Pateraki, M. Vakalellis, I. Spais\\
\thanks{Christos~Liaskos is with the Computer Science and Engineering Department,
University of Ioannina, Greece \& the Foundation for Research and
Technology Hellas (FORTH), Greece. E-mail: \protect\href{mailto:cliaskos@uoi.gr}{cliaskos@uoi.gr}.}\thanks{Ageliki~Tsioliaridou is with FORTH, Greece. E-mail: \protect\href{mailto:atsiolia@ics.forth.gr}{atsiolia@ics.forth.gr}.}\thanks{Konstantinos Georgopoulos, George Morianos and Sotiris~Ioannidis
are with the Technical University of Crete \& FORTH, Greece. E-mails:
kgeorgopoulos@mhl.tuc.gr, imorianos@isc.tuc.gr, sotiris@ece.tuc.gr.}\thanks{Iosif Salem, Dionyssios Manessis and Stefan Schmid are with the Technical
University of Berlin, Germany. E-mails: iosif.salem@inet.tu-berlin.de,
dionysios.manessis@tu-berlin.de, stefan.schmid@tu-berlin.de.}\thanks{Dimitrios Tyrovolas, Sotirios A. Tegos, Prodromos-Vasileios Mekikis,
Panayiotis D. Diamantoulakis, George K. Karagiannidis, Alexandros
Pitilakis and Nikolaos Kantartzis are with the Aristotle University
of Thessaloniki, Greece. E-mails: ${\text{\{tyrovolas,tegosoti,vmekikis,padiaman,alexpiti,kant,geokarag\}}}\text{@auth.gr}$.}\thanks{Anna Tasolamprou, Odysseas Tsilipakos and Maria Kafesaki are with
FORTH, Greece. E-mails: ${\text{\{atasolam,otsilipakos,kafesaki\}}}\text{@iesl.forth.gr}$.}\thanks{Ian~F.~Akyildiz is with Truva Inc., USA. E-mail: \protect\href{mailto:ian@truvainc.com}{ian@truvainc.com}.}\thanks{Andreas Pitsillides is with the University of Johannesburg (visiting
Professor), South Africa. E-mail: andreas.pitsillides@ucy.ac.cy.}\thanks{Maria Pateraki is with ORAMAVR, Switzerland. E-mail: maria@oramavr.com.}\thanks{Michael Vakalellis and Ilias Spais are with Aegis Research Ltd., Germany.
E-mails: ${\text{\{mvakalellis,hspais\}}}\text{@aegisresearch.eu}$.}}
\maketitle
\begin{abstract}
We present a new approach to Extended Reality (XR), denoted as iCOPYWAVES,
which seeks to offer naturally low-latency operation and cost effectiveness,
overcoming the critical scalability issues faced by existing solutions.
iCOPYWAVES is enabled by emerging PWEs, a recently proposed technology
in wireless communications. Empowered by intelligent (meta)surfaces,
PWEs transform the wave propagation phenomenon into a software-defined
process. We leverage PWEs to: i) create, and then ii) selectively
copy the scattered RF wavefront of an object from one location in
space to another, where a machine learning module, accelerated by
FPGAs, translates it to visual input for an XR headset using PWE-driven,
RF imaging principles (XR-RF). This makes for an XR system whose operation
is bounded in the physical-layer and, hence, has the prospects for
minimal end-to-end latency. Over large distances, RF-to-fiber/fiber-to-RF
is employed to provide intermediate connectivity. The paper provides
a tutorial on the iCOPYWAVES system architecture and workflow. A proof-of-concept
implementation via simulations is provided, demonstrating the reconstruction
of challenging objects in iCOPYWAVES-produced computer graphics.
\end{abstract}


\IEEEpeerreviewmaketitle{}

\section{Introduction}

Extended Reality (XR) is an emerging concept that includes spatial
computing technologies such as Augmented Reality (AR), Mixed Reality
(MR), and Virtual Reality (VR)~\cite{siriwardhana2021survey,akyildiz2022xr}.
Users with smart glasses, smart phones or head-mounted displays can
observe virtual content that does not exist in reality. XR will profoundly
change our lives across many areas, e.g., entertainment, manufacturing,
sports, and remote healthcare. For example, users with MR smart glasses,
e.g., Microsoft HoloLens, can share their real-time view with experts
and receive step-by-step remote assistance, which can significantly
improve worker productivity.

Motion-to-photon latency is a critical parameter in XR applications.
For example, when a VR gaming user presses a button on a controller,
the VR virtual content has to be rendered based on this motion. The
latency from motion to display should be lower than $20$~ms to avoid
motion sickness~\cite{wang2019multimodal,bermejo2021survey}. Currently,
the main contributor to latency in XR systems is the need for frequent
and successive crossings of all the Open Systems Interconnection (OSI)
model layers~\cite{wang2019multimodal,bermejo2021survey}. The information
from a multitude of sensors (cameras, lidars, sensors, actuators,
microphones) needs to be gathered in highly-confined timeslots to
a local server near a user. This information necessarily traverses
a network, is queued, and processed in very tight time windows, necessitating
high throughput wireless and wired networking infrastructure supporting
state of the art time-sensitive protocols, and high throughput computing
at the server side, commonly utilizing multiple, expensive CPUs and
GPUs. Depending on the XR targeted scale, the infrastructure requirements
are such that only large companies with datacenter infrastructure
near the end user can uphold them. Even at such cases, and apart from
the commitment of technological resources (capital expenses), the
operational expenses, such as the associated energy footprint, can
be exorbitant~\cite{wang2019multimodal,bermejo2021survey}. 
\begin{figure}[!t]
\centering{}\includegraphics[viewport=45bp 30bp 707bp 615bp,clip,width=1\columnwidth]{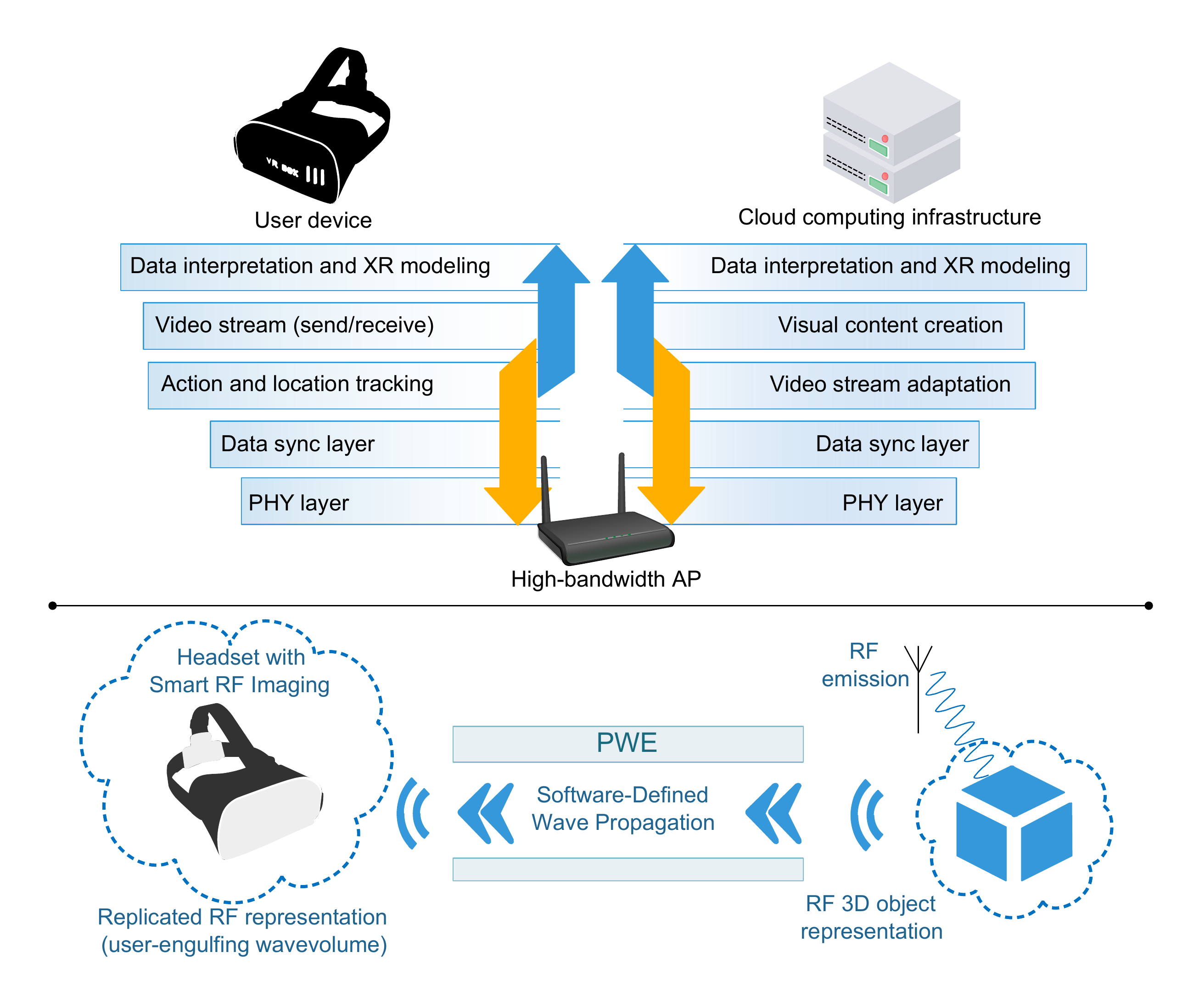}
\caption{{\small\textcolor{black}{High-level comparison between the existing
full-stack solutions for XR (top) and the proposed iCOPYWAVES approach
(bottom), whose operation remains bounded within the physical layer. }}}\label{fig:compareXR}
\end{figure}

XR is in urgent need of innovative solutions which provide low-latency
operation and cost-effectiveness. The present work seeks to explore
this path and meet the stringent performance requirements based on
two technological pillars, namely Radio Frequency (RF) imaging and
Programmable Wireless Environments (PWEs). RF imaging is a direction
stemming from physics, where RF waves are used for detecting the location
and shape of an object, as opposed to visible light imaging~\cite{RFimaging10,RFimaging20,RFimaging21pre,RFimaging98}.
Commonly, a single-frequency RF wave source emits waves upon a 3D
scene (much like the sunlight illuminates the objects around us),
and the scattered waves are collected by an array of receivers. The
captured wavefront is then mapped to the visual representation of
the object through analytical insights or machine learning approaches.
The second pillar of our proposal are PWEs, which constitute a recent
direction in wireless communications~\cite{PWE1,PWE2}, which is
already expected to be massively deployed in multiple environments
within 6G~\cite{ListofIS47:online}. PWEs transform the wireless
propagation phenomenon into a software-defined resource. PWEs are
created by coating all major surfaces in a space, such as walls and
ceilings in a floorplan, with programmable metasurfaces, also known
as software-defined metasurfaces (SDMs). In the broad sense, metasurfaces
are thin arrays of electromagnetically small elements, the so-called
meta-atoms, that are made tunable most commonly by incorporating tunable
impedance elements in general~\cite{RIS1,RIS2}. PIN diodes~\cite{srep06693,s41928-018-0190-1,s41928-021-00554-4,9020088,5184479,s41467-020-17808-y,9206044,9133266}
(common), MicroElectroMechanical Switches~\cite{MEMS} and ASICs
(more exotic)~\cite{ASIC1,ASIC2} (and others~~\cite{Mehdi2017990,Liu2019,He20161285,Shrekenhamer2013,Shen2011,Fan2019,Tasolamprou2019720})
are some popular circuit components employed to provide tunable impedance
and even more advanced capabilities. Metasurfaces are essentially
engineered materials that have customized and user-variable interaction
with impinging electromagnetic (EM) waves. Anomalous steering, absorption,
polarization control are exemplary manipulations performed by a metasurface
to impinging waves. Metasurfaces are created with an abundant array
of alternative cost-effective processes ranging, e.g., from standard
printed circuit boards (3D), to 3D printing and computerized numerical
control (CNC) milling~\cite{RISmanuf_3dp,RISmanuf_cnc,RISmanuf_pcb_rev}.
Moreover, PWEs abstract the underlying complex physics and allow the
tuning of a massive set of metasurfaces, inspired by the Software-Defined
Networking (SDN) paradigm to achieve operational logic-physics separation~\cite{SDNs,IoMbook}.

Based on these available components, the present study contributes
a new XR approach called iCOPYWAVES (\uline{i}ntelligent \uline{copy}ing
of RF \uline{wave}fronts/wavevolume\uline{s}), which simplifies the
XR architecture as shown in Fig.~\ref{fig:compareXR}. Through a
physical layer-bounded operation, iCOPYWAVES favors cost-effectiveness,
and ideally allows for nearly-speed-of-light end-to-end operation,
favoring scalability through naturally low-latency operation. The
core-idea is to\emph{ use PWEs to intelligently copy an RF wavefront,
or an RF wavevolume, from one location to another within a space.}
The RF wavevolume replication, i.e., copying a 3D EM field\emph{ }to
the location of an RF imaging device,\emph{ engulfing it within the
replicated field,} is of particular interest\emph{. }This would allow
the RF imaging device (envisioned to be embedded in a user's XR headset
in the future), to operate \emph{without the need for gyroscopes and
location tracking sensors}. As the user, e.g., rotates his/her head,
an RF imaging device embedded on his/her headset continuously reads
the corresponding part of the wavevolume, yielding the proper view
of the 3D object, even completely without assistance from sensory
devices or external computing elements: Machine learning-empowered
RF imaging reconstructs the 3D object using the copied wavefront as
input, and inserts it into an XR application setting in the proper
format. 

In summary, this paper presents a tutorial on the architecture and
workflow of iCOPYWAVES. Moreover, iCOPYWAVES is well-aligned with
forthcoming 6G infrastructure, which it reuses without further requirements.
Furthermore, the paper identifies and discusses challenges involved
in the end-to-end system implementation, covering all aspects, from
the advanced manipulation of EM waves provided by metasurfaces and
the PWE control algorithms necessary to implement wavefront copying,
to attaining efficient RF imaging reconstruction and insertion onto
an XR setting. Moreover, the proposed system advances the concept
of RF imaging to producing precise computer graphics for XR, i.e.,
XR-RF, as opposed to the coarse imaging capabilities of RF imaging
in the traditional use of the term~\cite{5.0076022,advs.201901913,j.patter.2020.100006,s41377-019-0209-z,1.4935941,srep23731}.
Promising results in this sense are also provided via a simulation-driven
implementation of the iCOPYWAVES system, which includes precise simulation
of the wireless propagation aspect and a machine learning component
trained to produce computer graphics in a challenging setup.

The remainder of this paper is organized as follows. Section~\ref{sec:-Background}
surveys the background work on XR systems, RF Imaging and PWEs. Section~\ref{sec:-The-iCOPYWAVES}
presents the iCOPYWAVES system architecture. Evaluation via simulations
follows in Section~\ref{sec:Evaluation}. Research challenges are
highlighted in Section~\ref{sec:Research-directions}, and the paper
is concluded in Section~\ref{sec:Conclusion}.

\section{State-of-the-art and limitations}\label{sec:-Background}

The great barrier that stands between current technology and remote
XR presence is the extremely stringent motion-to-photon latency, which
should not exceed 20~$m\text{sec}$ in order to avoid motion sickness
and enable lifelike experiences \cite{vr1,vr2}.

The motion-to-photon latency includes any delay incurred by motion
capture, encoding, communication, sensor fusion, processing, actuator
control, rendering and decoding of each frame. The different tasks
that lie on the critical path of the motion-to-photon latency, and
their associated timings include~\cite{vr3,vr4}: 
\begin{itemize}
\item Sensor sampling and synchronization: 1-5~$m\text{s}$ (high-end tracking). 
\item Scene rendering: 4-16~$m\text{s}$ (for 60Hz display). 
\item Display scanning: 2-16~$m\text{s}$ (60Hz, depending on the employed
technology). 
\item Photon emission: 1-2~$m\text{s}$ (depends on employed technology). 
\end{itemize}
These factors on their own can already violate the 20~$m\text{sec}$
total time frame. Moreover, the latency of the network that interconnects
multiple users is not even taken into consideration in these factors,
and can introduce additional latency in the range of $1-20$~$m\text{sec}$
on its own. (Notice that a part of this latency is unavoidable, and
stems from the finite speed of EM wave propagation. Thus, notice that
this range can be even overly optimistic for world-wide communication).

In AR, processes such as object identification, registration, or retrieval
of data already take considerable time~\cite{siriwardhana2021survey};
whereas, in VR, the sheer processing throughput required by the video
stream is highly taxing. On top of them, the display scanning and
photon launching further contribute to this delay. Even when considering
high-end hardware and processing techniques, these tasks will take
5 to 8~$m\text{sec}$. All in all, this leaves about 12 to 15~$m\text{sec}$
for the transmission, processing and reception of information, which
are the key enabling functionalities of remote XR presence.

The OSI layer crossing and the network latency combined is a barrier
that completely prevents the development of practical real-time remote
XR presence applications and will continue to be so unless it is reduced
at least tenfold. This constitutes an important shortcoming and reduces
the XR potential. Consequently, XR is currently a largely individual
experience, and at best allows participation of multiple locally connected
users. Bringing XR to the next level to enable lifelike interactive
and human-centric remote presence will require significant scientific
and technological advances.

In sharp contrast, iCOPYWAVES is a new approach to the XR concept.
With the combination of RF imaging and PWEs, the proposed system requires
no visual or gyroscopic sensory equipment, especially in the wavevolume
replication case. This means that data serialization and communication
via networking is required only in the remote site case, and can be
avoided even within, e.g., a large building or a mall. Moreover, even
in the networking case, the information that needs to be serialized
is simple waveforms that need not be understood or logically processed.
Instead, these wavefronts can be sampled and send over the wire directly,
or even undergo a physical-to-physical signal conversion with $\mu\text{sec}$
latency, such as wireless-to-optical and optical-to-wireless.

Here we also make note of existing network standards that also offer
minimal latency. Third Generation Partnership Project (3GPP) coined
the term ultra-reliable low-latency communications (URLLC)~\cite{URLLC},
which defines a target latency of 1~$m\text{sec}$ at a packet loss
ratio of $10^{-5}$ for 32-byte packets on the wireless radio access
network (RAN) segment~\cite{ran2018}. Another enabler is deterministic
networking (DetNet)~\cite{DetNet}, coined by Internet Engineering
Task Force (IETF), guaranteeing specific latency and jitter bounds
for packets routed through the core network segment. 

\subsection{Programmable Wireless Environments with Software-Defined Metasurfaces}\label{subsec:Rel2.1}

PWEs are end-to-end systems for controlling a wide array of metasurface
types~\cite{9109701,9171580,cryst11091089,Cui2014,Glybovski20161,dash2022active},
in order to apply deterministic control over the wireless propagation
process~\cite{PWE1}. A PWE is created by coating planar objects--such
as walls and ceilings in an indoor environment--with rectangular
and individually addressable metasurface panels with inter-networking
capabilities~\cite{end2endIRS,IoMbook}. The latter allows a central
server to connect to any metasurface unit, read its state and deploy
a new EM function in real-time and in a standards-compliant manner.
PWEs seek to provide a full protocol stack, clarifying the physical,
network, control and application layers of the complete system~\cite{end2endIRS}.
Moreover, PWEs leverage an SDN-inspired separation of concerns and~\cite{end2endIRS}: 
\begin{enumerate}
\item model the metasurface wave manipulation types (e.g., steer, split,
absorb, etc.) as software functions invokable via an application programming
interface. This makes the metasurface capabilities accessible to software
developers at large, without requirements for in-depth knowledge of
the underlying physics.
\item Maintain an abstracted, graph-based view of the system state, transforming
the PWE configuration optimization problem (i.e., how to tune each
SDM to serve a set of wireless devices), into a classic resource slicing
problem.
\end{enumerate}
Moreover, PWEs define the system workflow, from the discovery of a
PWE by a user device, to the statement of objectives and to its service,
in a generalized multi-tile, multi-use setting~\cite{PWE2019network}.
PWE as a generic control system goes beyond wireless communications,
exerting deterministic control over mechanical, acoustic and thermal
propagation~\cite{IoMbook,EMAPI}. Within the EM domain the PWE focus
is to craft EM vector field distributions, and not only reductions,
such as scalar power levels. To this end, PWEs treat metasurfaces
in their most generic way of operation, i.e., converters of surface
current distributions. Impinging waves create a surface distribution
\textquotedblleft A\textquotedblright{} upon an intelligent surface,
and embedded control elements convert it to a state \textquotedblleft B\textquotedblright{}
that yields the required EM field as a global response. This model
is denoted as software-defined metasurfaces (SDMs). Finally, PWEs
are created by massive deployments of SDMs in a space, covering all
major surfaces within it in a tiled sense, e.g., the ceiling and all
the walls in an indoors setting. The overall operation is typically
in the near-field, in the sense that PWE SDMs are not modeled as concentrated
at a single point in space~\cite{PWE2019network}.

Smart radio environments (SREs) constitute a concept that focuses
on the signal processing aspects of wireless communications, and especially
in conjunction with AI techniques~\cite{renzo2019smart}. The channel
control type is stochastic and the enabling technology are phase shifter
grids, which are commonly denoted as reflectarrays or intelligent
reflective surfaces (LIS, IRS or RIS)~\cite{PWE2,qian2021optimize}.
SREs typically assume very few RIS units, sparsely deployed within
a space, and in the far field in general. Based on these premises,
the goal is to iteratively optimize the phase shifter states (free
variables) in order to maximize a scalar quantity representing a wireless
communication objective (fitness function)~\cite{qian2021optimize}.
Additionally, given the theoretical signal processing focus of smart
radio environments, the required protocols, system workflows and integration-to-infrastructure
processes are commonly left undefined in the literature, i.e., inherently
assuming that an underlying PWE or related system stack or similar
is in place. In a layered sense, PWE is a top-to-bottom systemic approach,
while the smart radio environment is a layer-specific study (channel
modeling with RIS and applications).

PWEs and SREs are recent directions in long-standing but disparate
research efforts towards controlling the wireless propagation environment,
as opposed to the device end-points~\cite{iet-com.2010.0544,hist1,hist2,hist3}.
Approaches have explored the placement of passive reflectors to increase
coverage in a space~\cite{hist2}, to employing reflectarrays as
active alternatives~\cite{iet-com.2010.0544,hist1,hist3}. The SRE
direction consolidated the latter approach, and established the term
RIS to denote half-wavelength reflectarrays that are employed for
communication purposes~\cite{direnzo}. PWEs constituted a distinct
approach towards deterministic propagation control, and a departure
from the stochastic principles of preceding efforts. As such, the
uses of PWE go beyond communications and electromagnetism, exemplary
enabling deployments within high-precision imaging devices in order
to either increase their efficiency, or counter-balance manufacturing
imperfections~\cite{liaskos2021next}.

As such, PWEs and SREs have conceptual similarities, but also vast
differences regarding the capabilities and intended uses of each technology.
(Moreover, the terminology about SDMs/PWEs and RIS/SREs is still in
convergence, which can give rise to inaccurate classifications in
the literature). In this paper, we clarify that iCOPYWAVES requires
deterministic control over the EM propagation, as well as a full-stack
implementation in order to operate:

Specifically within the EM propagation control field, both SREs and
PWEs have been successful in mitigating path loss, fading and Doppler
effects at large~\cite{RIS1,Alamzadeh2021,9171580,huang2020holographic,renzo2019smart,PWE2},
albeit at different settings, overall costs and varying degrees of
efficiency. Moreover, the existing works treat device-to-device communications
only, which corresponds to a point-to-point wave replication. In contrast,
iCOPYWAVES studies the complete wavefront and wavevolume replication,
i.e., in 2D and 3D respectively, and its applications to XR imaging. 

Therefore, the remainder of this paper employs the terms PWE and SDMs. 

\subsection{From RF Imaging to XR-RF Imaging}

RF imaging is the general process of detecting attributes of hidden,
embedded or remote objects using RF waves (i.e., roughly within $300$~MHz
and $300$~GHz). RF Imaging can either strive for precision, attempting
to detect geometrical parameters such as shape, size and location
of an object, or for quantitative parameters, such as coarse composition
existence of features (such as cracks in a concrete slab), etc. Synthetic
aperture radars and ground-penetrating radars constitute some well-known
RF imaging approaches~\cite{RFimaging10}.

By leveraging metasurfaces, three-dimensional RF imaging beyond the
diffraction limit was made possible, with low-profile apertures, without
the need of lenses, moving parts or phase shifters, reducing the cost,
size, complexity and power demands of conventional imaging technology~\cite{Hunt2013310,Hunt:14,Wan2017}.
Tunable millimetre metasurfaces have been so far implemented to produce
spatially diverse patterns in the vast microwave and low THz spectrum~\cite{8477051,8718481,Hosseininejad2019,Diebold20181529}.
Approaches have striven for task-specific and low latency RF Imaging,
i.e., employing metasurfaces to illuminate scenes in a targeted effort
to look for particular objects from a given set of possible templates~\cite{5.0076022,advs.201901913,j.patter.2020.100006}.
The low-latency aspect has been studied without taking into account
the presence of a protocol stack, such as PWE, which is necessary
to provide standards-compliant interconnectivity. Task-specificity
in RF illumination can also be considered as the next step from random
illumination~\cite{1.4935941,srep23731}. The common denominator
of these state of the art approaches in RF-Imaging is the coarse quality
of the end-outcome. Crisp graphics required for XR is not a possibility
due to the inherent limitation of the emploed technological assumptions,
essentially when employed in conjunction with RIS/SREs for the reasons
detailed in Section~\ref{subsec:Rel2.1}.

Moreover, it is a well-known fact that the resolution of RF imaging
in its simplest form, i.e., \textit{\emph{i) employing a single RF
source, ii) based on the amplitude of the received signals only, and
iii) using analytical formulas to perform the image reconstruction}}\emph{,
}\textit{\emph{is defined by the employed wavelength}}. Under such
conditions, high-resolution RF imaging would require high RF frequencies,
e.g., at least $30$~GHz. Interferometric synthetic-aperture radar
approaches can mitigate this limitation by employing phase and amplitude
RF processing techniques, operating within the $5\to15$~GHz range~\cite{Davis2008}.
For instance, it has been shown that the achievable resolution in
a reverberant environment as opposed to free space is orders of magnitude
better because the reverberation provides a generalized interferometric
sensitivity. This point was also proved via experiments with a RIS/SRE
targeted at localization~\cite{PhysRevLett.127.043903}. 

In differentiation, the PWEs employed in the iCOPYWAVES are also used
for accurately manipulating the RF wavefront scattered by a 3D object
(i.e., apart from its replication), and essentially act as spatially
distributed synthetic-aperture radar system. In other words, via PWEs
the received wavefronts are optimized for imaging, while unwanted
effects (such as sidelobes) are canceled out or driven away from the
RF receiver. Moreover, the wavefront replication approach means that
a high-quality RF receiving device can be potentially shared for XR
imaging tasks. Regarding the relation of iCOPYWAVES to task-specific
illumination, the proposed XR approach is intended to be generic and
objective towards the visualized scene, and not task-specific, which
could potentially lead to biased XR visualization. However, task specificity
can be attained by programming the PWE accordingly (e.g., copying
targeted aspects of a wavefront), or by using machine learning modules
specifically trained for the task in mind. Moreover, the paper shows
that machine learning approaches can learn to perform a wavefront
processing that is more complex than analytical formulas, which are
inherently limited by human or physical intuition. As a result, the
evaluation example included in the present study operated efficiently
\textit{\emph{at just $5$ GHz}}. 

Therefore, the proposed approach can promote a new direction for precise
computer graphics produced by RF imaging, that can be used in XR systems,
as opposed to the existing coarse RF imaging solutions. We denote
this research goal as XR-RF imaging and employ it in the remainder
of the paper. (In addition, it is noted that the present study bears
no similarity to the field of holographic RIS/SRE~\cite{gao2018nonlinear},
which refers to metasurfaces targeting precise electromagnetic control
in their near field, and is irrelevant to XR, despite the similarity
in their naming).

\section{ The iCOPYWAVES approach to XR}\label{sec:-The-iCOPYWAVES}

iCOPYWAVES seeks to provide the necessary infrastructure for: i) creating
RF wavefront representations of 3D objects, and ii) manipulating these
RF wavefronts with the ease of a \textquotedblleft copy-paste\textquotedblright{}
functionality, thereby virtually transferring an RF representation
of the original objects (their external structure and, perhaps, their
internal composition as well) to remote locations. 

\paragraph{System setup}

\begin{figure*}[t]
\centering{}\includegraphics[width=1\textwidth,viewport=0bp 0bp 796bp 635bp]{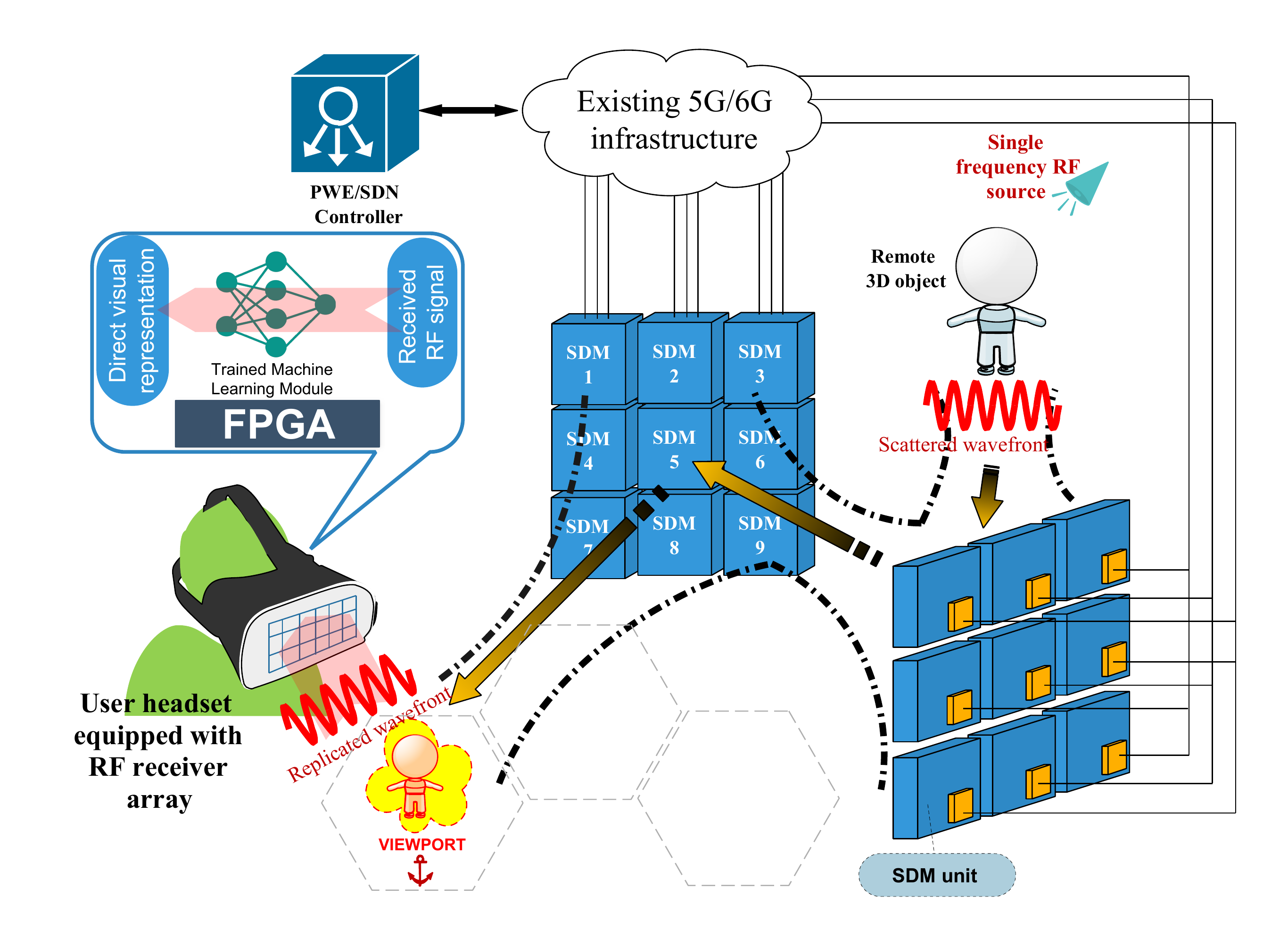}
\caption{{\small\textcolor{black}{The iCOPYWAVES end-to-end system workflow,
overviewing the steps from the creation of an RF scattered wavefront
from a remote object, to its visual representation at the user's headset.}}}\label{fig:PWEworkflow}
\end{figure*}

We firstly describe the overall workflow of the iCOPYWAVES system
in its operation phase, assuming the setup shown in Fig.~\ref{fig:PWEworkflow}.
The setup uses a PWE comprising of a set of SDM, connected to an SDN
controller for orchestrating them towards achieving a software-defined
electromagnetic wave propagation. We note that the SDM and the SDN
controller are considered to be parts of existing 5G/6G communications
infrastructure. iCOPYWAVES simply adds another use of the same infrastructure
for XR.

We proceed to describe the general end-to-end workflow of iCOPYWAVES.
A new user enters the system, and associates himself with the existing
PWE. (The reader is redirected to related studies for details on the
user registration to the PWE, and the PWE deployment and initialization~\cite{end2endIRS}).
The new user is equipped with an iCOPYWAVES-compatible headset, which
incorporates a MIMO antenna array, and an FPGA (Field-Programmable
Gate Array) hosting a pre-trained machine learning component. 
\begin{rem}
In the scope of the present paper, Generative Adversarial Networks
(GAN) are being used exclusively, but without loss in generality.
\end{rem}
The GAN is trained to translate wavefronts impinging upon the MIMO
antenna array into a visual XR outcome, such as a 3D cloud point,
or Left/Right eye video streams (cf. Section~\ref{subsec:GAN Training}). 
\begin{rem}
The actual video format produced by the GAN can be adapted freely,
to facilitate its direct integration to the underlying rendering workflow.
For instance, setting the output format to be a cloudpoint ensures
the 3D coherence of the reconstructed object. The choice of left/right
video streams can be simpler to integrate in the XR experience, but
may require extra processing to ensure coherence. In such cases, the
extra processing power is also assumed to be offloaded to the FPGA.
\end{rem}
In the meantime, a remote object receives impinging waves emitted
by an RF source. In the setup shown in Fig.~\ref{fig:PWEworkflow},
the source is a cheap, single-frequency signal generator connected
to a horn antenna. The waves scatter upon the remote 3D object, and
arrive at SDM units around it. Then, it is the task of the PWE to
copy the scattered wavefront around the vicinity of the XR user's
headset, essentially engulfing the headset within the replicated wavefront. 
\begin{rem}
The headset operates in a closed local loop, continuously translating
the received RF wavefronts into visual outcomes, and without further
communication with the SDN server. An additional benefit is that the
user's head rotations are automatically translated to the corresponding
changes in the visual outcome.
\end{rem}
The exact way of how the iCOPYWAVES operates to serve the user XR
objectives may be adapted to the availability of a new user coarse
localization system:

\textit{If a user localization system does not exist}: The SDN controller
sets up viewpoints where remote users are replicated. The new user
must then walk and get closer to the replicated wavefront, which is
much like what he/she would do for a real object. No information on
the user's location or mobility is required. The PWE configuration
is viewed as a large optimization problem~\cite{zheng2019optimize,zhao2020optimize,wu2020optimize,abeywickrama2020optimize,qian2021optimize},
and ensures that the multi-user viewpoints do not interfere. This
can be accomplished with the wavefront routing logic shown in Fig.~\ref{fig:PWEworkflow}.
In other words, the SDN controller solves the PWE configuration problem
by finding air-routes that are disjoint from one another.

\textit{If a user localization system exists}, then the SDN controller
also knows the approximate location of the user device. This enables
the controller to perform a versatile and fast \emph{rerouting of
the EM wavefronts/wavevolumes}~\cite{liaskos2020TCOM,PWE1,PWE2019network,ITU_JFET},
thus following the user's mobility pattern. This can allow for more
immersive XR experiences, where, e.g., a remote avatar follows the
user around, much like a tag-along character in a video-game. Notice
that knowledge of the user's position does not include the orientation
of his/her head, but rather only his/her position on the floorplan
in a relatively coarse X-Y basis, as shown in Fig.~\ref{fig:PWEworkflow}.
The new user remains immersed in the replicated RF wavevolume. Thus,
his/her head movements and minor X-Y dislocations are automatically
translated to corresponding wavefront readings, allowing the machine
learning module to produce the updated object view automatically.

Note that:
\begin{itemize}
\item Fig.~\ref{fig:PWEworkflow} illustrates an indoors operation. However,
this is not at all\label{Fig.-illustrates-an} restrictive. Remote
operation across any distance is discussed later in two variations,
in Section~\ref{subsec:Pilot-3:-Copying}, including the transfer
of other signal types, such as audio and haptic.
\item RF waves do not carry coloring information. (They carry, however,
material composition information which opens interesting new applications
discussed later, in Section~\ref{subsec:Pilot-3:-Copying}). Smart
coloring can be offloaded to the GAN (i.e., by training the GNA to
artificially color objects, as demonstrated in Section~\ref{sec:Evaluation}),
or be based on extra rendering steps, employing a single color photo
of the 3D object. In both cases, the computational overhead is assumed
to be offloaded to the FPGA.
\end{itemize}
We proceed to detail the various operation phases of the iCOPYWAVES
system.

\subsection{The iCOPYWAVES GAN training phase}\label{subsec:GAN Training}

\begin{figure*}[!t]
\begin{centering}
\includegraphics[width=1\textwidth]{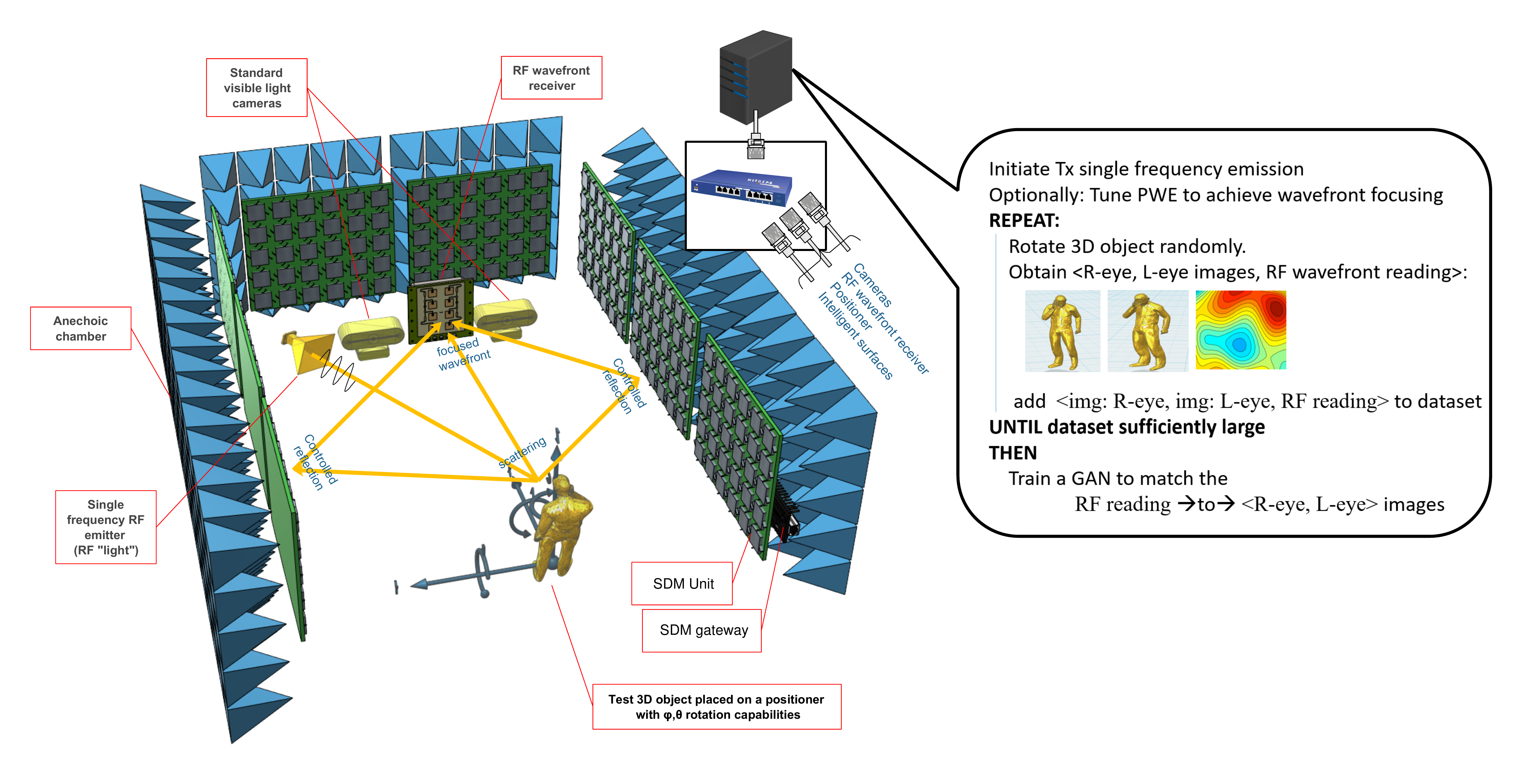}
\par\end{centering}
\caption{{\small Overview of the envisioned iCOPYWAVES GAN training phase (which
occurs offline), showcasing the successful translation of RF readings
into visual representations of the corresponding 3D objects.}}\label{fig:Pilot1}
\end{figure*}
We consider the setup of Fig.~\ref{fig:Pilot1}, which consists of
a test 3D object, a positioner (i.e., a device that can programmatically
rotate an object in spherical coordinates), a set of standard visible
light cameras, an RF wavefront receiver, a single frequency RF emitter,
and a set of SDM units. All devices are connected to a central SDN
controller and they are--optionally--located in a controlled electromagnetic
environment, i.e., an anechoic chamber that mimics operation in free
space~\cite{ITU_JFET}. 

The RF emitter acts as the ``RF light'' that illuminates the 3D
object with EM waves. The 3D object scatters the waves, which are
collected by the RF wavefront receiver. The SDM units assist by focusing
more scattered waves on the RF wavefront receiver. The SDN controller
collects the RF wavefront reading, as well as regular, visible light
photos of the 3D object from a set of cameras. The process repeats
as follows: 
\begin{itemize}
\item The positioner rotates the 3D object to a random rotation. 
\item A data structure of $\left\langle \text{RF reading},\left\{ \text{photos}\right\} \right\rangle $
is obtained and is added to a data set. 
\end{itemize}
The process repeats until a sufficiently large number of data set
entries has been obtained, in order to train a GAN to produce the
photos from a given RF reading~\cite{GANpix2pix,GANpix2pix_mri}.
The photos can then be used to reconstruct the 3D object into an XR
setting.
\begin{rem}
The anechoic chamber shown in Fig.~\ref{fig:Pilot1} and onward is
applicable to the GAN training phase, where accurate EM propagation
control via the PWE is desirable. In real conditions, the GAN should
be trained with artificial noise and interference patterns in a controlled
environment first, before final deployment. We note that this is a
measure of reducing the \textquotedblleft free variables\textquotedblright{}
at this stage of the iCOPYWAVES presentation. However, this does not
preclude future GAN approaches that will be able to get trained in
a drop-in fashion in real and uncontrollable conditions.
\end{rem}

\subsection{The iCOPYWAVES operation phase I: Copying wavefronts/wavevolumes
indoors}\label{subsec:Pilot-2:-Copying}

\begin{figure*}[!t]
\begin{centering}
\includegraphics[width=1\textwidth]{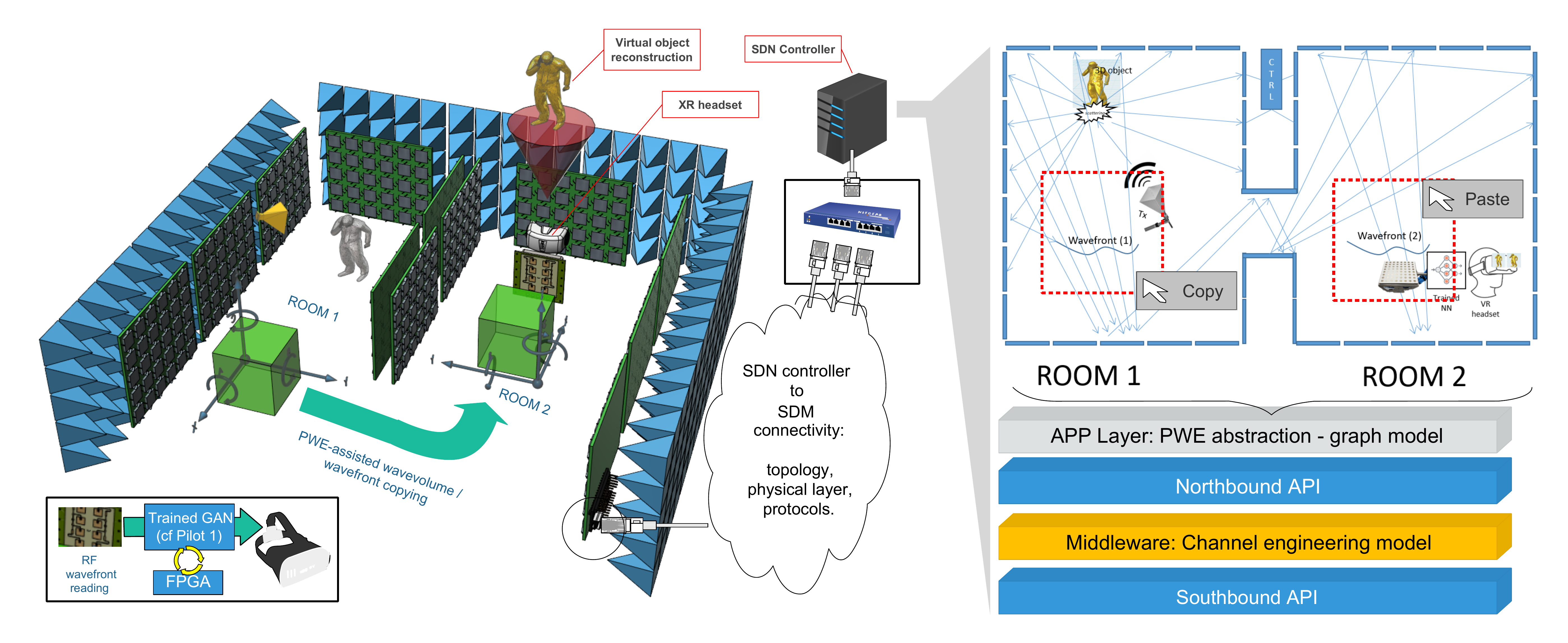} 
\par\end{centering}
\caption{{\small Overview of one envisioned iCOPYWAVES operational phase, showcasing
the copying of a wavefront carrying XR visualization in an indoors
environment. }{\small\textit{\emph{While the left visualization illustrates
the controlled environment of an anechoic chamber, the phase is intended
to operate as-is in regular environments.}}}}\label{fig:Pilot2}
\end{figure*}

The first operation phase that we study, focuses on indoors settings,
considering the setup of Fig.~\ref{fig:Pilot2}. (We assume that
the outcomes of the training phase have been produced, and a trained
GAN exists, which is able to reproduce a visual presentation of the
3D object from a corresponding RF wavefront).

The studied phase intends to provide the facilities for copying a
wavevolume from one location to another on in proximity, without the
need for over-the-wire data transmission. Therefore, the scenario
of this phase intends to demonstrate a case where a virtual 3D object
needs to be projected in relatively close vicinity to the original
object, e.g., within a building.

In the studied phase we consider a space separated into two compartments,
namely ``room 1'' and ``room 2'', using SDM placed on the walls.
The test 3D object in located in room~1, and is illuminated by the
``RF light''. The PWE is then configured to replicate the scattered
wavevolume from room~1 to room~2, where a receiving system exists.
The receiving system consists of an RF wavefront reader collocated
with the trained GAN, and an XR headset.

We note the following additional research tasks pertaining to the
studied phase: 
\begin{enumerate}
\item The trained GAN outputs are assumed to be pre-processed as needed
for feed into the XR system. 
\item The trained GAN/XR preprocessing system will be ported to an FPGA
in order to minimize the overall processing time. 
\item The SDM units comprising the PWE will be connected to an SDN controller.
The choice of the corresponding networking architecture, physical
means, protocols and topology may be optimized for low-latency and
low-cost. However, this assumed infrastructure can be part of the
existing 5G/6G communications infrastructure, and not be deployed
for the iCOPYWAVES system.
\item The PWE control is exerted by a protocol stack implemented within
an open SDN controller, thereby offering compatibility with the existing
SDN ecosystem. The stack comprises~\cite{IoMbook}: 
\begin{enumerate}
\item A northbound API that models the macroscopic behavior of the SDM units
in the form of a library of callbacks. Examples include the definition
of $\texttt{STEER()}$, $\texttt{SPLIT()}$, $\texttt{ABSORB()}$,
$\texttt{PHASE\_ALTER()}$ function etc., to describe the interaction
of an intelligent surface with an impinging EM wave. It is noted that
each SDM comes with a \emph{configuration }codebook, supplied during
its quality check right after manufacturing. This codebook matches
each possible callback to the corresponding configuration, i.e., the
collective states of the SDM embedded control elements (e.g., PIN
diodes). 
\item A middleware in the form of a channel engineering model capturing
the XR-RF imaging case. Notably, macrscopic PWE callbacks come with
unintended microscopic side-effects. This can include unintended sidelobes
during EM wave steering, as well as fading phenomena arising from
SDM location imprecisions and imperfect control over the EM waves
in general. The middleware provides a channel engineering model which
will receive as inputs a set of activated northbound API callbacks,
and will return as output the resulting, precise channel behavior. 
\item A southbound API that provides connectivity compatibility between
the SDN controller and the PWE control network. 
\item A PWE abstraction layer, implemented on top of the northbound API,
that provides a macroscopic model of the complete PWE system in the
form of a graph. The objective is to provide a framework for decomposing
high-level wave-copy commands into smaller, tractable problems, such
as finding sets of paths within a graph that comply with a given set
of criteria. Thus, a wavefront/wavevolume copy command will be treated
as a set of point-to-point EM wave routing decisions. 
\end{enumerate}
\end{enumerate}
Moreover, we stress that the XR workflow remains bounded within the
physical layer, from the creating of a wavefront scattered from a
3D object, to its FPGA-driven translation to visual information.

\subsection{The iCOPYWAVES operation phase II: Copying wavefronts/wavevolumes
remotely over the wire}\label{subsec:Pilot-3:-Copying}

\begin{figure*}[!t]
\begin{centering}
\includegraphics[clip,width=1\textwidth,viewport=20bp 20bp 800bp 380bp]{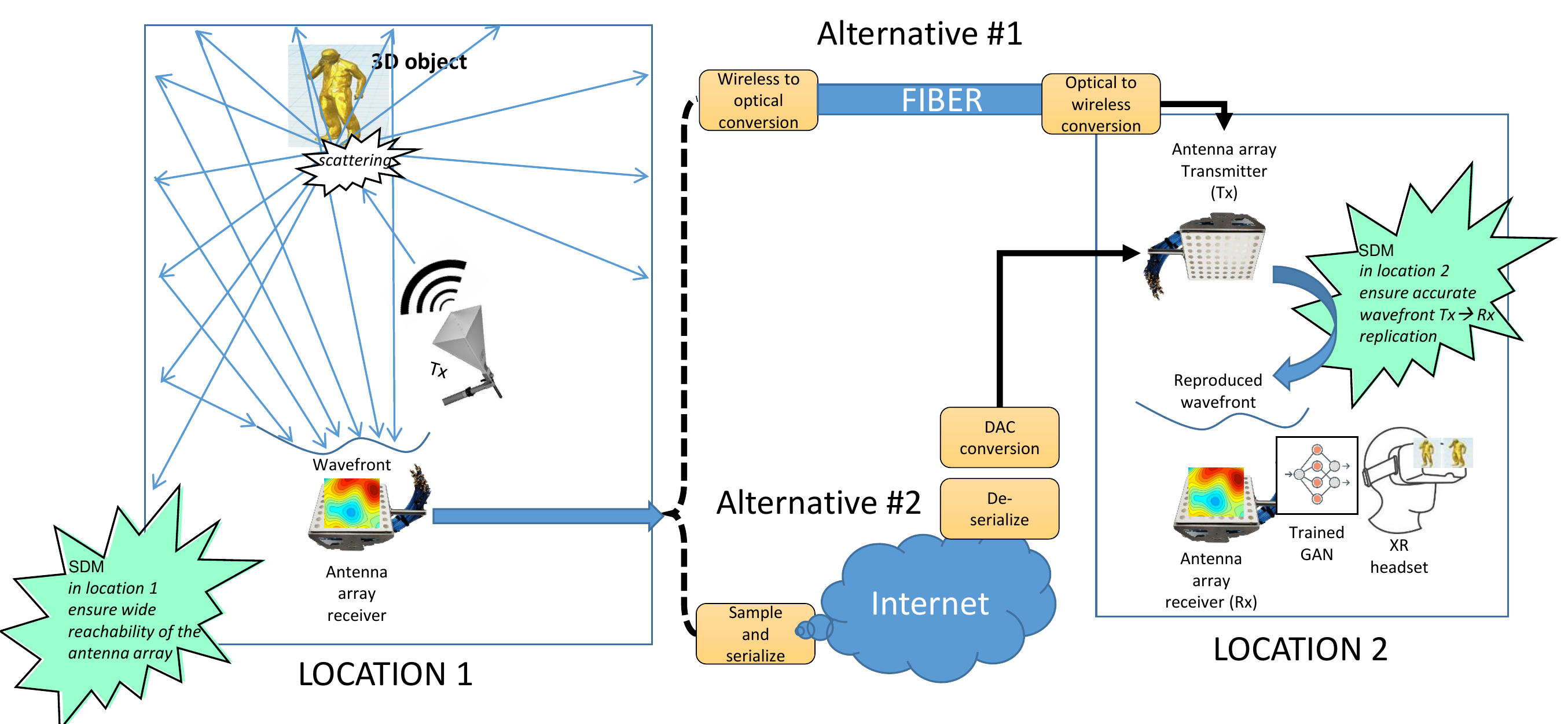}
\par\end{centering}
\caption{{\small Overview of the }{\small\uline{first}}{\small{} approach}{\small\textbf{
}}{\small for remote operation. This scenario requires a receiving
antenna array in location 1, and a transmitting antenna array in the
remote location 2. The goal is to transfer the wavefront impinging
over the receiving antenna array in location 1 over direct fiber (wireless-to-optical
happens at the physical layer with practically zero latency). In lack
of fiber availability, the Internet can be used as a common alternative.
Within location 2, the PWE handles the wavefront replication from
the Tx to the Rx (collocated with the user's XR headset). This approach
requires regular SDMs, }{\small\uline{without}}{\small{} wavefront sensing
capabilities.} Additionally, we note that this approach is also the
default way for carrying all complementary information, such us sound
and any sensor (e.g. haptic) feedback.}\label{fig:Pilot3-simple}
\end{figure*}

\begin{figure*}[!t]
\begin{centering}
\includegraphics[width=1\textwidth]{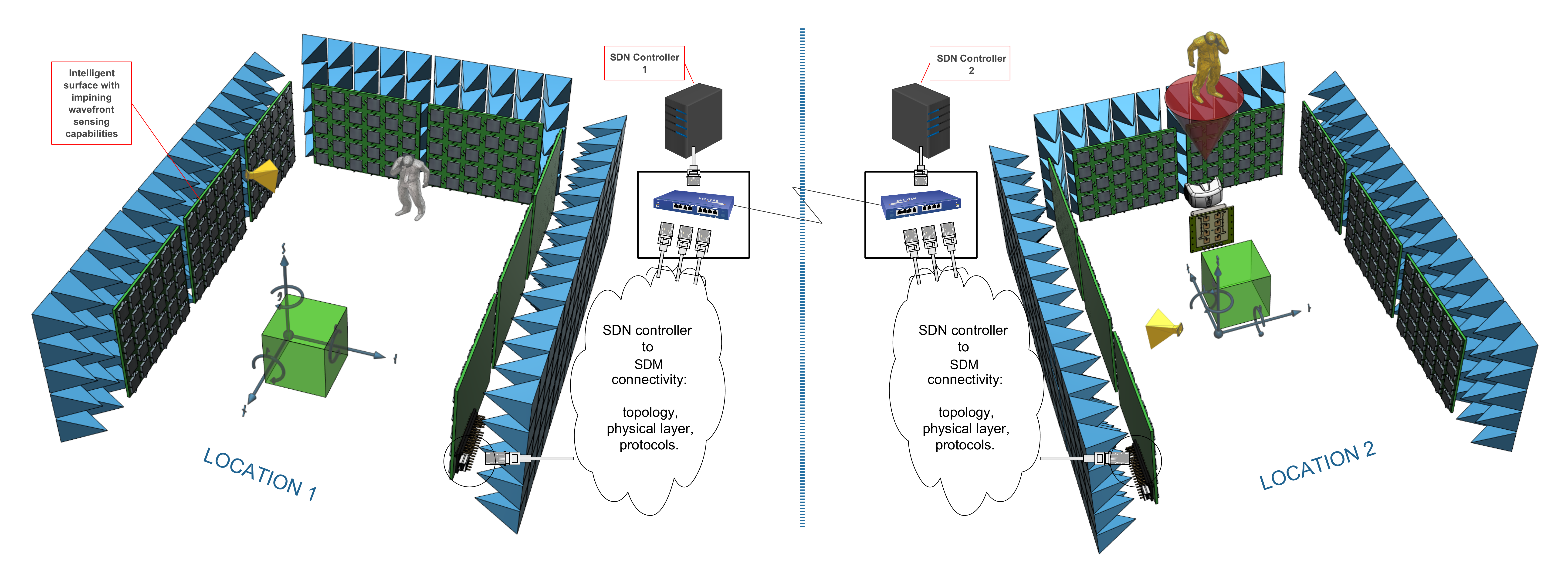}
\par\end{centering}
\begin{centering}
\includegraphics[clip,width=1\textwidth,viewport=0bp 0bp 960bp 460bp]{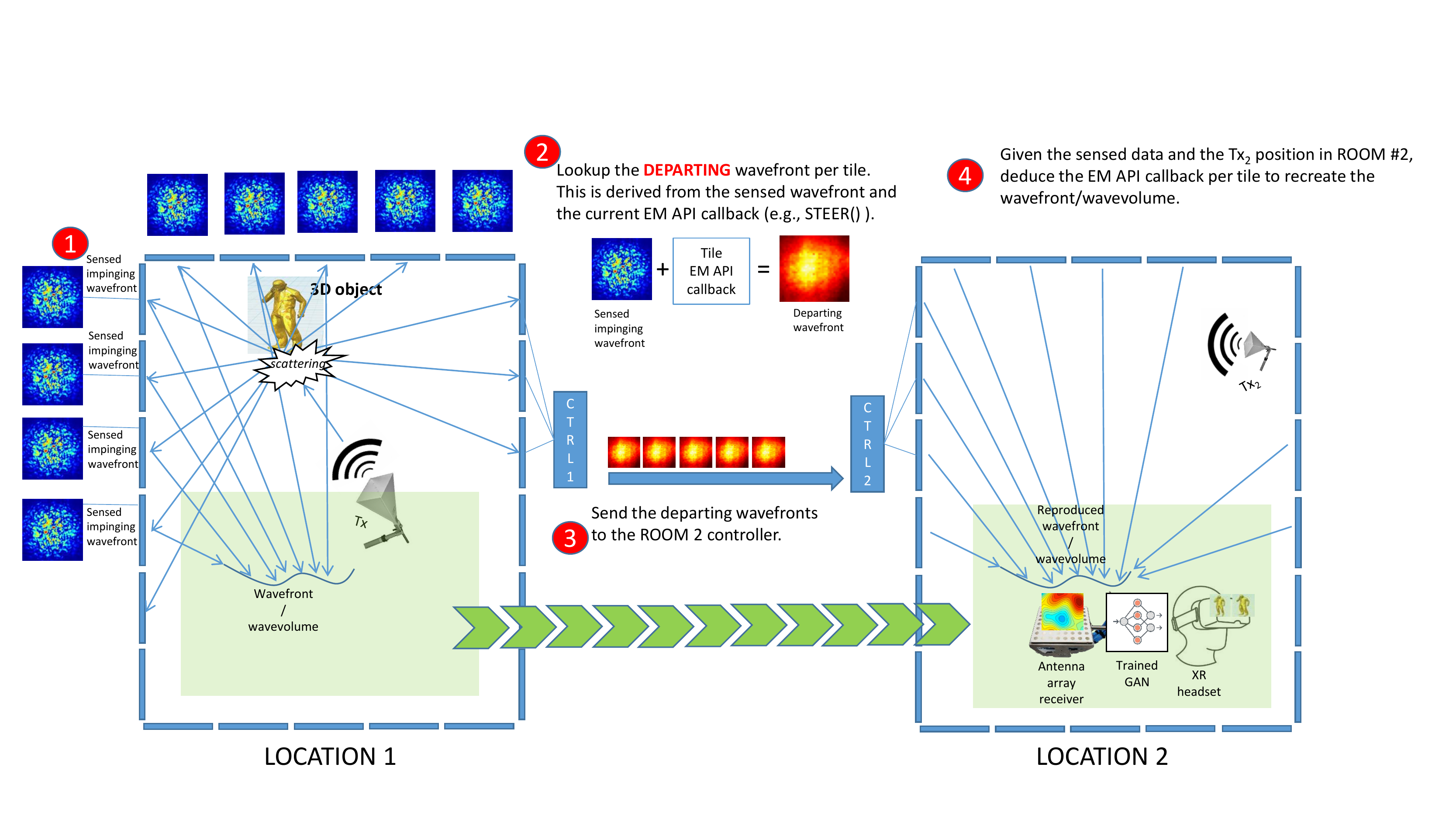}
\par\end{centering}
\caption{{\small Overview of the }{\small\uline{second}}{\small{} approach}{\small\textbf{
}}{\small for remote operation. (Top) Showcasing the copying of a wavefront
carrying XR visualization in an indoors environment. This approach
does not require extra antenna arrays, while it is able to replicate
wavevolumes as well as wavefronts. (Bottom) Workflow for the \textquotedblleft over-the-wire\textquotedblright{}
reproduction of the intended wavefront/wavevolume. This approach requires
advanced SDMs, }{\small\uline{with}}{\small{} wavefront sensing capabilitites.}}\label{fig:Pilot3-complex}
\end{figure*}

We proceed to describe the remote operation of iCOPYWAVES in Figures~\ref{fig:Pilot3-simple}
and~\ref{fig:Pilot3-complex}. This phase extends the previous case
with operation ``over-the-wire'', in order to serve any remote location(s).
As shown in Figures~\ref{fig:Pilot3-simple} and~\ref{fig:Pilot3-complex},
we consider two separate locations, each with its own PWE controller.
Location 1 contains the actual 3D object to be remotely projected
at another location (2) containing the XR receiving system.

Interconnectivity is accomplished by taking advantage of the same
SDN controller employed in phase~I. Essentially, interconnecting
the two controllers over the Internet establishes a pathway for altering
the PWE of, e.g., location~2 based on input from location~1. Nonetheless,
given that there is no actual means for direct wave propagation between
the two locations, we require a means on converting the wavevolume
of location~1 into a format that can be transferred over the wire
interconnecting the two locations.

To this end, we consider two approaches for phase~II:

\textbf{Approach \#1,} shown in Fig.~\ref{fig:Pilot3-simple}, remains
cost-effective while maintaining the low-latency operation prospects.
Moreover, it focuses explicitly of wavefront replication by using
two extra antenna arrays. The first antenna array is placed anywhere
in location~1. The PWE of location~1 is tasked with focusing the
scattered waves from the 3D object over this antenna array. The second
antenna array is placed anywhere within location~2, and it acts as
a wavefront transmitter. The PWE in location~2 is tasked with replicating
the wave emission from this transmitting antenna array to the vicinity
of the user. Notice that, as shown in Fig.~\ref{fig:PWEworkflow},
the copied wavefront can act as a viewport. It can remain anchored
to a specific place, and any user standing in front of it can see
the remote object.

The wavefront read from the antenna array of location~1 can then
be sampled, serialized and send to the transmitting antenna array
of location~2. This can occur over the Internet, subject to the (uncontrollable)
latency that this alternative entails. Another alternative is to employ
direct RF-to-optical / optical-to-RF converters, as allowed by the
available infrastructure~\cite{RF2optical}. In this approach, every
element of the receiving antenna array is mapped directly to a optical
light variations over a dedicated wavelength and travels at the speed
of light along a fiber. The conversion itself is of direct, physical-to-physical
nature, and has trivial latency.
\begin{rem}
\textit{\emph{We note that this approach--as a general concept--is
employable across all phases to transfer sound (and haptic per case)
signals. Such signals can be generated as wireless analog waveforms
(e.g., FM in case of sound), and undergo a similar wireless-to-optical/optical-to-wireless
workflow.}}
\end{rem}
\textbf{Approach \#2} follows a different, more exploratory, research-natured
premise. It does not require the extra antenna array pair of approach~1,
but introduces SDMs with impinging wavefront sensing capabilities
at location~1. Moreover, it targets the remote wavevolume replication,
albeit with a latency trade-off. In overview, for the second approach: 
\begin{itemize}
\item Location~1 is coated with intelligent surfaces that can sense the
impinging wavefronts upon them. (The coating can be partial, and this
decision is subject to research considerations). 
\item The impinging wavefront over each surface, along with the corresponding
active callback, is sent directly to the SDN controller at location~2.
Either SDN controller~1 or~2 can use this information to deduce
the wavefront that \emph{departs} from each intelligent surface at
location~1. 
\item The SDN controller then calculates the SDM callbacks for each surface
at location~2 that create the same departing wavefronts as in location~1.
(If the two rooms/locations are not identical, the creation can refer
to an appropriate distance away from each surface, rather than over
the surface itself). 
\end{itemize}
The SDN controller calculations will be sped up via FPGAs to minimize
the overall latency. 

\paragraph{Further prospects. }

iCOPYWAVES ideally replaces the common XR workflow consisting of the
discrete steps, i.e., 3D scanning $\rightarrow$ sensory data gathering
$\rightarrow$ system synchronization $\rightarrow$ rendering $\rightarrow$
projection, with physical layer-bounded operation operating as 
\begin{multline*}
\text{\textbf{RF\,Imaging\,\&\,EM\,wave\,guiding\,(speed\,of\,light)}}\rightarrow\\
\left[\text{\textbf{FPGA-driven\,GAN(\ensuremath{\mu}sec)}}\right]\rightarrow\text{\textbf{projection}}
\end{multline*}

\noindent Note that the wide time margin expected to be gained by
the iCOPYWAVES approach enables hybrid approaches as well, where: 
\begin{itemize}
\item Virtually any rendering process can precede projection. This is particularly
important, given that \textit{RF signals cannot see colors}. Therefore,
the proposed system provides the 3D geometry of the object, while
the XR application rendering can proceed to perform smart texturing
and coloring (e.g., from a simple photo taken by a local camera once
per few seconds, in a hybrid approach). Moreover, as shown in Fig.~\ref{fig:toy_example},
the GAN can also learn to color 3D scenes \textit{quite effectively
and in tandem with the XR-RF imaging process}, providing a good range
of (even complementary) options, without any camera involved during
operation. 
\item The internal material composition of the 3D object can also be visualized
in real-time, since \textit{RF signals can penetrate objects}. This
is a unique capability of the proposed approach, especially useful
to medical imaging and industrial material telemetry XR applications~\cite{de2020survey,bermejo2021survey},
compared to existing XR systems. 
\end{itemize}
Finally, remote sites can be flexibly interconnected with over the
Internet with some simple, direct sampling-serialization-deserialization
of the EM waves. The same approach is employed for carrying complementary
data, such as sound and haptic data collected by sensors, \textit{\emph{cost-effectively
and at near-light speed}}.

\textbf{Regarding security}, we note that the system outlined in Fig.~\ref{fig:PWEworkflow}
offers capabilities for robust user access control and privacy. \textit{Firstly},
the outline user authorization process, performed via the SDN controller,
ensures the regular degree of security that contemporary information
systems can offer. \textit{Second}ly, the PWE control offers exquisite
control over the route taken by the replicated wavefronts (cf. Fig.~\ref{fig:PWEworkflow}),
ensuring that they avoid unauthorized users in real-time~\cite{PWE1}.
\textit{Finally}, PWEs have the capability to locally scramble and
then descramble a traveling wavefront around a user~\cite{liaskos2019adhoc}.
This capability can also be used as follows. As shown in Fig.~\ref{fig:toy_example}
a received wavefront can be viewed as a colored map (using any arbitrary
representation process). Therefore, upon a new user entering the system,
i) the PWE can be instructed to scramble his received wavefront(s),
while ii) the user receives upon entering the system unique descrambling
instructions than can restore the intended wavefront colour map to
its intended form prior to the GAN processing. In contrast, unauthorized
users will receive the same wavefront as ``white noise''.

\section{Proof-of-concept Evaluation}\label{sec:Evaluation}

\begin{figure*}[!t]
\begin{centering}
\includegraphics[width=1\textwidth]{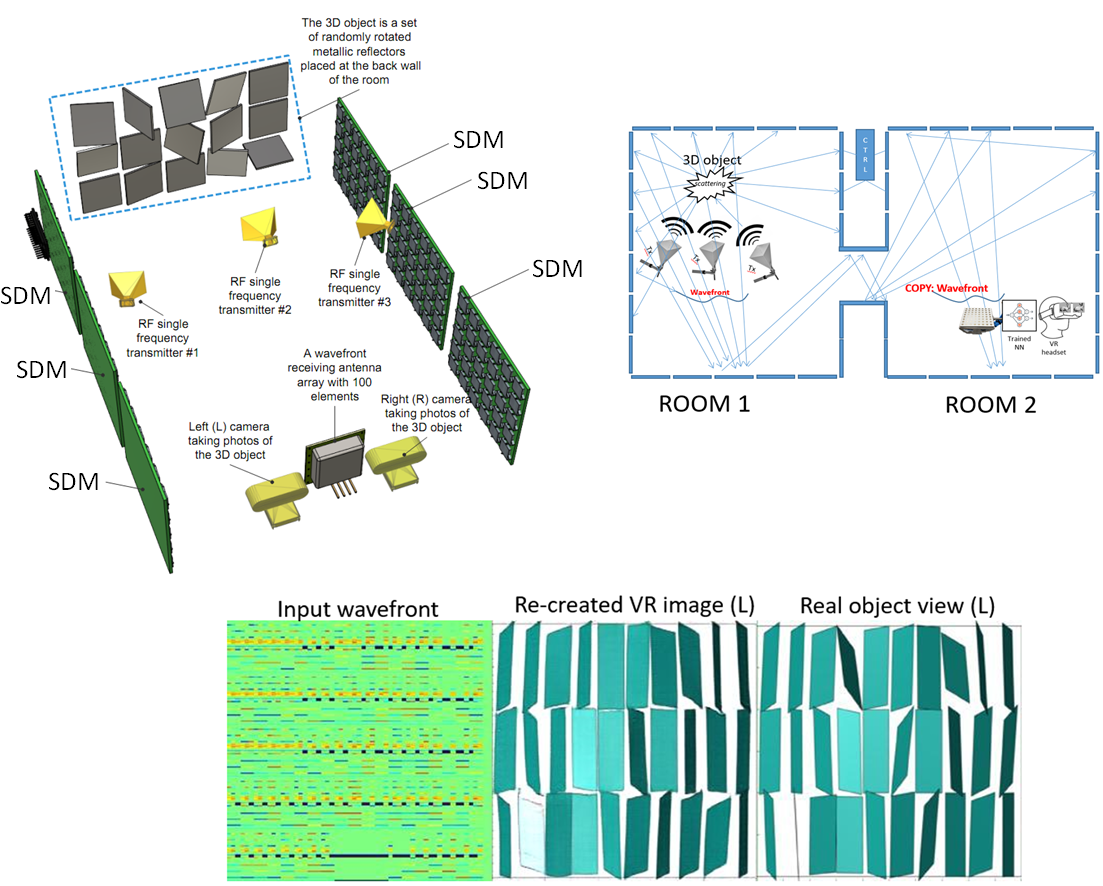}
\par\end{centering}
\caption{{\small Validation setup: an XR-RF imaging process in a simple room
(top left), where the 3D object is a set of randomly rotated rectangular
reflectors placed on a wall. Three RF transmitters (yellow horn antennas)
emit 5~$G$Hz waves upon the 3D object which scatter around it, and
an antenna array with 100 elements gets a corresponding reading. Two
standard cameras--(L)eft and (R)ight--take visual snapshots of the
3D object. The process is repeated 1000 times, each time rotating
the 3D object randomly, thus, creating a dataset for training a GAN~\cite{GANpix2pix},
which then translates an RF reading to a visual outcome. Once the
training is complete, a second room is added to the layout with all
walls covered with metasurfaces (top right). The original wavefront
is copied from room 1 to another place in room 2. The trained GAN
recreates the visual snapshot (bottom).}}\label{fig:toy_example}
\end{figure*}

A proof-of-concept scenario, validating the PWE-enabled XR operation,
is simulated in realistic RF ray-tracing software~\cite{mededjovic2012wireless},
and as shown in Fig.~\ref{fig:toy_example}. 

\begin{figure}[!t]
\begin{centering}
\includegraphics[viewport=135bp 280bp 480bp 500bp,clip,width=1\columnwidth]{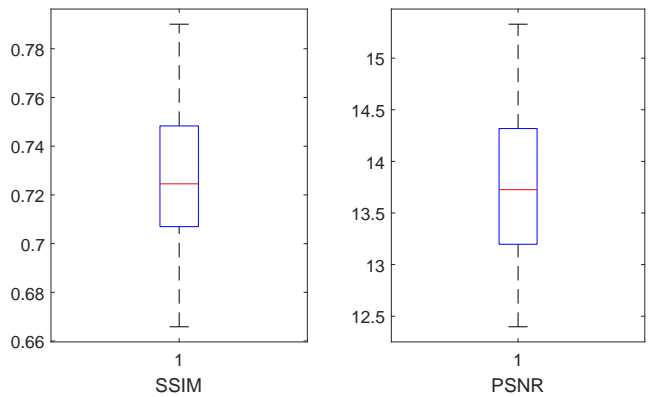}
\par\end{centering}
\caption{Boxplots of the attained SSIM and PSNR image comparison values over
$100$ object reconstructions via iCOPYWAVES.}\label{fig:boxplots}
\end{figure}

\begin{figure}[!t]
\begin{centering}
\includegraphics[viewport=220bp 140bp 400bp 660bp,clip,width=0.8\columnwidth]{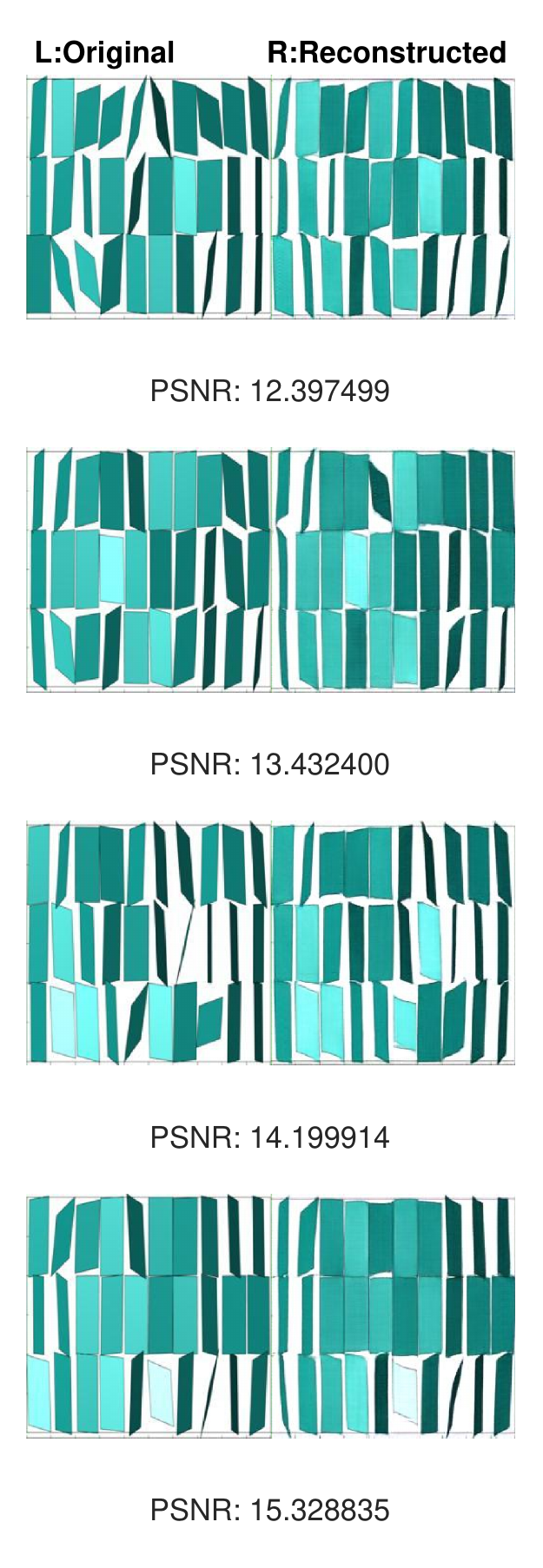}
\par\end{centering}
\caption{Indicative iCOPYWAVES outcomes spanning the corresponding PSNR boxplot
range of Fig.~\ref{fig:boxplots}}\label{fig:PSNR}
\end{figure}

\begin{figure}[!t]
\begin{centering}
\includegraphics[viewport=220bp 140bp 400bp 660bp,clip,width=0.8\columnwidth]{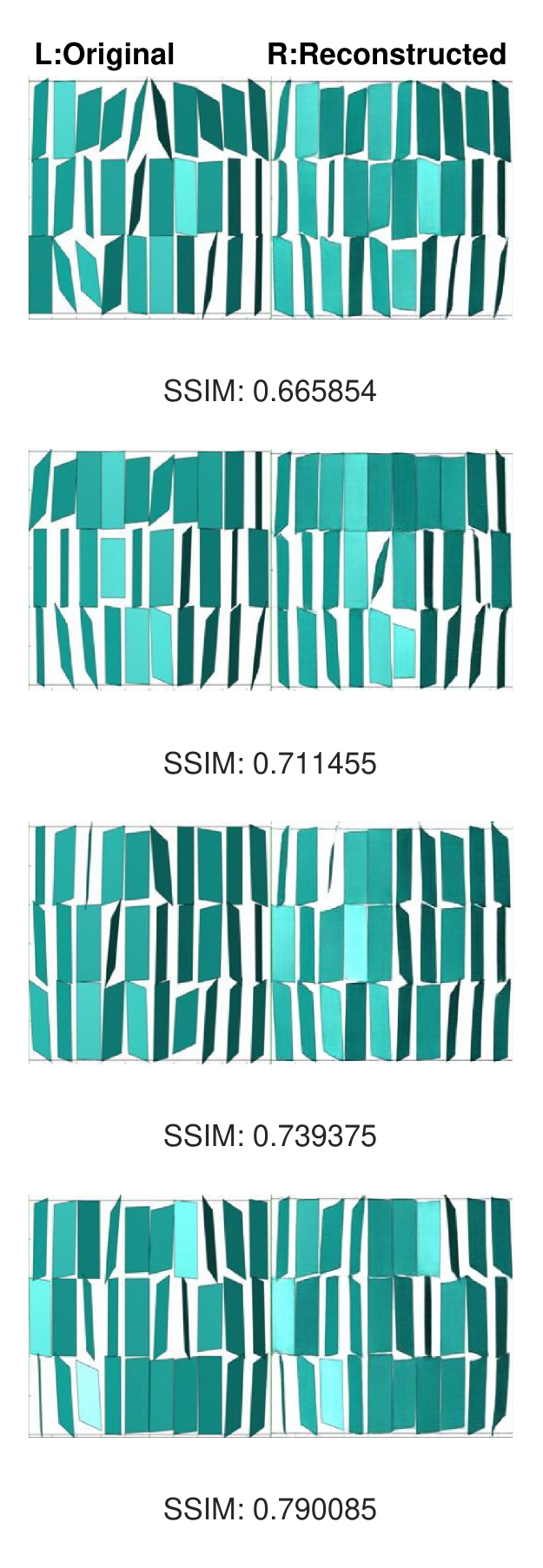}
\par\end{centering}
\caption{Indicative iCOPYWAVES outcomes spanning the corresponding SSIM boxplot
range of Fig.~\ref{fig:boxplots}}\label{fig:SSIM}
\end{figure}

Here, we assume an XR-RF imaging process in a simple room, where the
3D object is a set of randomly rotated rectangular reflectors placed
on a wall. Three RF transmitters (yellow horn antennas) emit 5~$G$Hz
waves upon the 3D object which scatter around it, and an antenna array
with 100 elements gets a corresponding reading. Two standard cameras--(L)eft
and (R)ight--take visual snapshots of the 3D object. The process
is repeated 1000 times, each time rotating the 3D object randomly,
thus, creating a dataset for training a GAN~\cite{GANpix2pix}, which
then translates an RF reading to a visual outcome. Once the training
is complete, a second room is added to the layout with all walls covered
with metasurfaces (top right). A PWE with two adjacent rooms attempts
to copy the wavefront scattered from a random 3D object (arbitrarily
rotated rectangular metallic reflectors) from ROOM 1 to ROOM 2. A
pre-trained Generative Adversarial Network (GAN) recreates the image
of the 3D object. The original wavefront is copied from room~1 to
another place in room~2, using a PWE optimization engine~\cite{PWE2019network},
and the trained GAN recreates the visual snapshot. We compare the
real and reconstructed images via image comparison techniques, and
particularly via the Peak SNR (PSNR) and the Structured Similarity
Indexing (SSIM) methods~\cite{Sara2019-bk}. Boxplots of the attained
values are given in Fig.~\ref{fig:boxplots} per method. Moreover,
indicative graphics corresponding to the ranged of quantized comparison
outcomes are given in Fig.~\ref{fig:PSNR} and~\ref{fig:SSIM}.

The outcome of this case study already implies that the XR system
can efficient provide copying wavefronts/wavevolumes. In particular
we observe that:
\begin{itemize}
\item The simulated system yields an output close to the ideal one, while
also performing artificial coloring of the generated image, also in
good agreement with the expectation. Notice that a similar GAN has
been used successfully in a medical setting, for translating positron
emission tomography (PET) scans to magnetic resonance imaging (MRI)~\cite{GANpix2pix_mri}. 
\item The 3D object (randomly rotated rectangular metallic reflectors) constitutes
a very challenging case, given that this object has no coherent structure.
If the object had coherence (e.g., if the system targeted specific
object types, like humans, robotic arms, furniture, etc.), then a
``forward correction'' could have been performed as a final part
of the GAN structure to provide a more accurate output. Another promising
approach would be to employ a GNeRF type of GAN~\cite{meng2021gnerf},
which requires multiple RF readings, but i) directly yields the 3D
object as output, and ii) offers noise resilience owed to the assembly
of the multiple RF readings. 
\end{itemize}
We note that this example does not incorporate a real XR application
or an actual SDN controller, while also operating on the premise of
ideally performing metasurfaces.

The target functionality is based on the emerging technology of SDM
and PWEs developed in recent years~\cite{IoMbook,VISORSURF}. In
particular: 
\begin{itemize}
\item With regards to the metasurface part we assume a unique SDM design
and prototype operating at 5~$G$Hz, covering the complete process:
from EM analysis, to PCB design, to integration of electronics, and
to final prototype assembly~\cite{RIS1}. 
\item Additionally, we assume first of its class software family to interact
with metasurfaces in a physics-agnostic manner~\cite{EMAPI}. \emph{Firstly},
the EM wave manipulation types and their parameters are organized
in the form of a software library of callbacks, denoted as the intelligent
surface \emph{application programming interface} (API). This allows
computer networking experts to invoke the intelligent surface functionalities
without specialized knowledge in physics.\textbf{ }\emph{Second}ly,
the translation of such software callbacks into embedded circuit states
in real-time was achieved via the offline creation of a lookup database
(DB), via extensive measurements and simulations performed during
the intelligent surface manufacturing phase. This enabled the real-time
operation and re-adaptation of PWEs, while also accounting for unintended
effects (e.g., sidelobes), and imprecision stemming from factors such
as unintended directions of wave arrivals over SDMs.
\begin{figure}[!t]
\begin{centering}
\includegraphics[width=0.95\columnwidth]{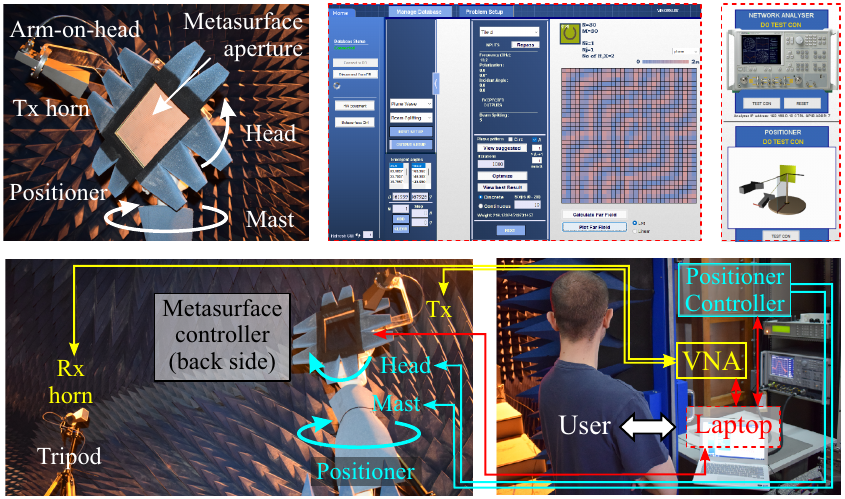}
\par\end{centering}
\caption{The contributed testbed for realizing an XR copying wavefronts/wavevolumes
system. Clockwise, from top left: Close-up photo of a metasurface
loaded on an automated positioner and illuminated by a fixed source;
screenshots from the developed software that calculates and assigns
configurations to a metasurface, and directs the measurement setup.}\label{fig:testbedPrototype}
\end{figure}
\end{itemize}
Finally, a complete implementation can take advantage of a completely
automated testbed combining SDM prototypes, dynamically positioned
transmitters/receivers, RF measurement devices, and an implementation
of the aforementioned software, all controlled via a central computer
(shown in Fig.~\ref{fig:testbedPrototype}-top)~\cite{ITU_JFET}.
The implemented testbed consists of a bistatic setup inside an anechoic
chamber, configured for the measurement of 3D scattering patterns.
The testbed employs an automated positioner with a rotating mast and
head; a metasurface and a transmitter (Tx) antenna are mounted on
the rotating head while a receiving (Rx) antenna is placed on a tripod;
the forward transmission of the system is measured with the help of
a vector network analyzer (VNA); The entire testbed is controlled
by the aforementioned software running on a laptop from the outside
of the anechoic chamber; essentially, the software must connect to
the VNA and the positioner mast (and head), for the acquisition of
a 2D or 3D scattering pattern, and to the SDM for passing different
commands.

\section{Research directions }\label{sec:Research-directions}

We proceed to discuss the research challenges towards the implementation
of the various components required for iCOPYWAVES system.

\subsection*{Portable hardware for XR-RF Imaging based on machine learning}

A crucial and integral part of the complete system is that of the
XR headset, which is going to incorporate a technology that ensures
real-time visual representation of electromagnetic stimuli from the
original room at a minimal energy-consumption footprint.

The purpose is to process the information arriving from the source
at a very high speed and the concept is to do so by using a GAN. (Other
promising machine learning approaches can be evaluated, posing a separate
but secondary challenge). On that topic, FPGA technology has proven
to be a highly attractive candidate for such tasks and applications.
This is owed to the fact that FPGAs can facilitate parallelized execution,
which is suitable for implementing neural networks such as the ones
employed within the context of GANs. FPGAs contain hardware resources
that can be designed and structured in a multitude of ways in order
to implement all sorts of architectures. Hence, FPGAs can also accommodate
machine learning inference models, offering the potential for full
parallelization at the same-layer neuron execution while accelerating
intra-layer execution by pipelining the machine learning computations.

Moreover, FPGAs offer a quantity of hardware resources and this leads
to the ability to select the optimum tradeoff between scaling and
performance, i.e. given a target throughput, the designer can work
towards the dedication of the corresponding hardware resources that
will lead to the satisfaction of that particular throughput specification.
Overall, the application specifications will pose a set of target
specifications such as minimum throughput, maximum area overhead and
maximum power consumption, and FPGAs offer the ability to investigate
different architectural implementations that yield the best results.

On top of that, it is possible to customize designs to the precise
arithmetic required by the neural network and, in fact, design for
the particular ones required by the different layers of the network.
For instance, it could be that, overall, a given neural network requires
INT8 arithmetic to complete. This however, does not mean that the
INT8 requirement is posed by all of the network's layers. It is possible,
therefore, with FPGAs to attribute the INT8 arithmetic requirement,
to the particular layer that requires it and design the other layers
with less demanding requirements such as INT2 and INT4. This has the
inherent potential for power saving through hardware minimisation.

Finally, it must be noted that FPGAs come in a wide variety of packages
with a highly diverse range of characteristics (e.g., thermal dissipation,
power consumption) making them suitable for cloud as well as edge
implementations. Hence, they fit seamlessly to the present concept
that requires the integration of FPGA-processing into a XR headset.
The headset will contain a custom packaging that will host the FPGA,
catering for proper thermal management. This board will contain all
the necessary electronics that will facilitate the FPGA utilisation
such as the conditioning circuitry, which will ensure that the electromagnetic
signals received by a MIMO antenna are appropriate for introducing
them to the FPGA GAN model. 

\subsection*{XR platform for XR-RF imaging and XR merge}

This challenge refers to the translation of the outputs of the XR-RF
imaging system (GAN output signals) to an immersive XR setting, providing
the necessary software platform for developing applications such as
teleconferences and education. 

The mitigation pathway for this challenge is twofold, one more experimental
and one following a more common norm, which also acts as a failover
strategy. The first approach is to translate the GAN outputs directly
into a video stream that is merged with the XR visual stream, with
as little preprocessing as possible. This approach has the potential
of a highly simplified operational workflow, resulting into a simplified
XR headset hardware in the future as well. 

The second approach follows a model-driven approach. Having an array
of possible object templates, it uses the GAN outputs to deduce the
best matching choice, ensuring that a coherent and artifact-free object
is finally rendered. An additional experimental direction is to take
into account the XR-RF imaging capability in providing insights about
the internal composition and material structure of a remote 3D object,
enabling applications such as remote medical imaging or inaccessible
material telemetry. 

\subsection*{Wireless channel engineering optimization for XR-RF imaging}

A major challenge is the microscopic wireless channel modeling, deducing
its behavior while copying wavefronts and wavevolumes. This challenge
requires a mature electromagnetic design and characterization process
as a prerequisite, building upon the previously described challenge,
and extending it as follows. 

\paragraph*{Topology}

In far-field imaging, EM waves impinge as plane waves, thus a small
difference in distance does not affect significantly the amplitude
of echo signals resulting from the distance between each point of
the target object and the \textquotedblleft radar\textquotedblright{}
being approximately the same \cite{buzzi}. Since passive localization
is utilized, it is important to evaluate as many as possible signal
routes from the transmitter to each receiver through the monitored
object. A distributed or a cascaded architecture can be utilized to
cover the space with either transceivers or SDMs to guarantee the
tracking of each path. Possible overlaps between coverage areas of
the APs can be resolved based on the angle of arrival measurements
when multiple antenna elements are available \cite{emma}. To control
the transmission and to reach beyond possible blockages a number of
SDMs has to be deployed in the physical space. In this case, ideally,
the whole indoor environment should be covered with SDMs in order
to achieve full control and nearly-deterministic sensing, however,
with simulations, an ideal number of SDMs can be extracted that accomplishes
an acceptable performance in terms of imaging of the indoor environment.

\paragraph*{Development of Appropriate Channel Models}

The SDM channel modelling accounts for a variety of possible EM/SDM
interaction types, including beam steering, beamforming, focusing,
modulation, and joint modulation and encoding with the transmitter
\cite{tegos},~\cite{PWE1}. Depending on the specific application
and wave transformation applied, the path loss and fading models will
be extracted via physics-level simulations and theoretically. (Approaches
exist for SRE/RIS systems as well, e.g.~\cite{9856592}. For the
PWE/SDM case, i.e., the designated approach for iCOPYWAVES, there
exist open physics simulation platforms that enable this approach
by the scientific community in general~\cite{CNSM}). An all-electromagnetic
architecture that prescribes PWE-user interaction in the radiative
near field should be followed. To this end, it is necessary to rely
upon physics-based models for the propagation of EM fields in the
proximity of metasurfaces and/or extract circuit models for the problem
formulation \cite{direnzo}. Accurate path-loss models for link budget
analysis, as well as fading models for sub-wavelength structures,
both at the microscopic level, should be developed based on the extension
of mathematical physics methods that capture the electromagnetic properties
of wave propagation in complex environments \cite{green}. Eventually,
the developed models will allow for a quick and scalable study of
the PWE control impact on the XR-RF imaging system, without the need
for time-consuming precise simulations. Subsequently, the optimized
algorithms are integrated for the implementation of the SDN controller
and the PWE control algorithms \cite{end2endIRS}.

\paragraph*{Dynamic Imaging Controller}

In smart XR-RF imaging, RF waves are emitted toward the objects under
examination to detect their structure \cite{robotic}. One or a set
of transmitters emit waves upon a 3D scene, and the scattered waves
are collected by an array of programmable metasurfaces. Depending
on the electromagnetic function applied at these metasurfaces by the
image controller, such as beam-steering, diffusion, etc., it is possible
to dynamically adjust the captured wavefront which is subsequently
mapped to the visual representation of the object through analytical
expressions and recent advances in machine learning \cite{tegos},~\cite{PWE1}.
The image controller should be able to adjust the scattering diagrams,
in order to capture instantly the differences in moving objects. Therefore,
it is imperative to explore the combination of these functions that
can be applied to the programmable metasurfaces and investigate their
effects on detection accuracy.

\paragraph*{Quantification of Replication Accuracy}

Regarding the impact of the wireless channel on the object detection
accuracy, multi-carrier waveforms with advanced peak-to-average power
ratio (PAPR) reduction techniques and single-carrier waveforms with
advanced modulation techniques are promising research directions,
offering reduced PAPR without sacrificing the spectrum efficiency.
Moreover, the design of waveforms for narrowband scenarios, characterized
by low processing complexity, could be explored to enable power saving.
Also, robustness to the restrictions posed by low-cost hardware, e.g.,
time and frequency offsets, should be examined. For example, on-off
modulation is simple to implement and can be useful if the target
data and access rate meet the specific scenario requirements. In addition,
the geometry of the environment can affect the accuracy of object
detection and replication. Complex geometries need high resolution
to detect fine details. High resolution can only be obtained with
high bandwidth, which would increase the cost of the total system.
In this direction, SDMs can assist in creating LoS links increasing
the resolution. Finally, predicting accurately the behaviour of the
physical assets is another challenge. To address this issue, advanced
AI tools can be utilized, which are capable of learning general and
complex patterns of the environment by exploiting historical information
and exploring future actions and decisions as well. Similarly, AI-optimized
constellations and demodulators can assist in decreasing PAPR.

\paragraph*{Waveform design}

Waveform design can start by implementing known signal forms, such
as chirp, and by utilizing reinforcement learning (RL) methods the
waveforms can evolve to adapt to each environment \cite{petropulu},
\cite{waveradar}. The key challenge in this research direction is
that sensing should be performed simultaneously with communication,
which facilitates the efficient utilization of so valuable system
resources such as bandwidth and power. Waveform design through AI
is an area of interest in the research community lately, as AI can
design unconventional waveforms that better handle the deteriorations
of the channel. In this case, reflectivity and scattering of materials
should be showcased \cite{liaskos1}. At first, supervised learning
could be implemented to evaluate the use of AI for ISAC and then a
complete RL framework can be implemented where no prior learning or
knowledge of the environment is necessary. The complexity of such
a system would be higher, but once a good model of the physical space
is created, little changes to the waveforms will be required based
only on the mobility and the dynamic changes in the environment.

\subsection*{PWE controller implementation and system component networking}

This challenge pertains to the modeling of the PWE at a macroscopic
level, designing algorithms for the orchestration of multiple SDMs
for wavefront/wavevolume copying, as detailed recently~\cite{end2endIRS}.
It integrates the outcomes of the channel engineering challenge, and
seeks to create the iCOPYWAVES SDN controller implementation. 

Towards this end, this challenge needs to account for the network
architecture interconnecting the controller to the SDM in the same
or remote locations, from the latency perspective. As intermediate
steps, the challenge can be tackled by SDM studying a PWE graph model
to abstract the underlying physics at an algorithmic development level.
Then, it can proceed to produce resource orchestration algorithms
based on this graph model, targeting the creation of wireless propagation
paths that perform the wavefront and wavevolume replication from one
location to another. Fault tolerance must also be studied, e.g., in
terms of system hardware imperfections, system component synchronization
irregularities, and user mobility. 

In parallel, meeting this challenge also entails taking into account
the component integration, i.e., the experimentally verified wireless
channel models, as well as the networking infrastructure required
to interconnect the PWE system controller to the SDM units. The implementation
of the PWE controller culminates this challenge, which completes the
infrastructure that iCOPYWAVES requires, apart from the user headset.

\subsection*{Design and manufacturing of SDM units for XR imaging}

Regarding the SDM technology required for iCOPYWAVES, challenges concern
the design, development and low-level characterization at the physical
layer. 
\begin{figure}[!t]
\begin{centering}
\includegraphics[width=0.95\columnwidth]{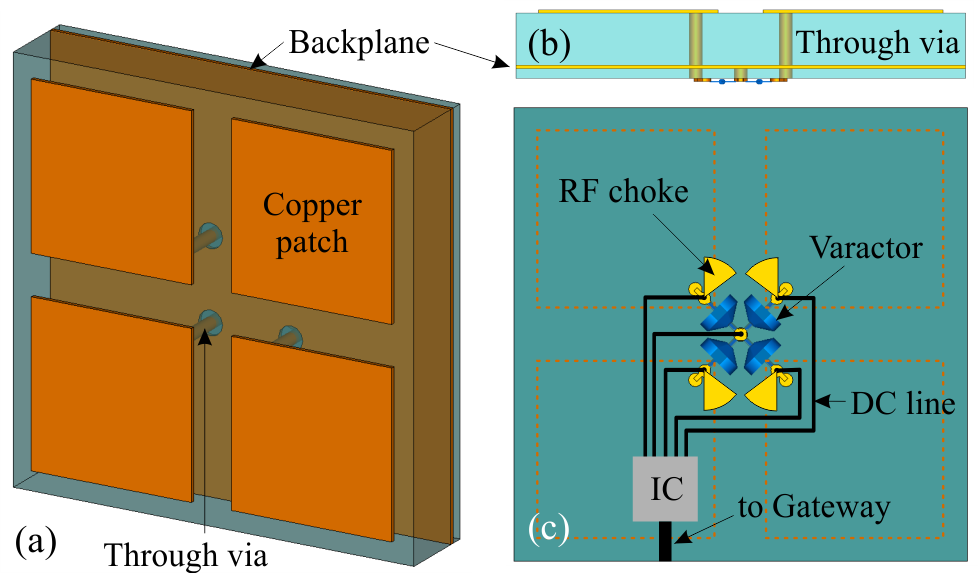}
\par\end{centering}
\caption{(a) Perspective top-side view of a unit cell composed of a 2x2 array
of square patches; this cell topology allows independent control of
reflection amplitude and phase in both polarizations. (b) Side view
of the unit cell, showing the through vias connecting the patches
at the top side to the actuators embedded in the bottom side. (c)
Bottom side view of the unit cell, where the actuators here correspond
to varactors, i.e., DC-voltage controlled variable capacitors; an
IC embedded in the cell receives commands from the gateway and applies
voltages to each of the four varactors effectively setting the cell
response; note that microwave engineering is required here, e.g.,
RF chokes to isolate the DC/RF lines. }\label{fig:testbedPrototype-1}
\end{figure}

This initially entails (a) the design of the SDM that will efficiently
perform the advanced wave control required for XR-RF imaging within
the PWE; (b) the electronic design of the control circuits and components
for the embedded SDM control, and (c) the advanced manufacturing of
patterned PCBs as SDM. 

Regarding the electromagnetic design of the SDM, we rely on solid
physical conclusions derived in our previous works~\cite{Liu2019,RIS1,9171580}.
More specifically, to be able to perform arbitrary wavefront manipulation
with the SDM units, we require local control over the complex surface
impedance of the metasurface. This requires individually-controlled
unit cells equipped with voltage-driven electronic actuators that
control both the reactive and resistive part of the surface impedance.
In the GHz regime, such actuators can be implemented with MOSFETs
configured as varactor (variable capacitance) and varistor (variable
resistor) elements. To be able to control the two linear polarizations
separately (and thus additionally perform arbitrary wave, it is also
desirable to electrically rotate the \textquotedblleft principal axes\textquotedblright{}
of the unit cell; this can be achieved by a composite unit cell design,
e.g., based on four patches (see Fig.~\ref{fig:testbedPrototype-1}),
where individual patches can be connected vertically, horizontally,
or diagonally~\cite{ITU_JFET}. 

The fabrication requirements of the proposed SDMs involve the capability
of manufacturing high-quality multiple-metallization boards using
high-frequency, low-loss specialty substrates. In addition, we need
high-aspect ratio through vias and blind vias with uniform metal plating
and good electrical conductivity characteristics in order to minimize
the resistive loss experienced by the loop currents excited in the
unit cell~\cite{pitilakis2022multi}. Finally, the assembly of the
controller chips is based on closely-space solder balls to enable
the dense packing of unit-cells/controllers. The feasibility of all
the above enabling capabilities has been demonstrated in earlier work~\cite{manessis2020manufacturing}. 

Following the electromagnetic/electronic co-design of the PCBs/SDMs
and the manufacturing of the designed PCB/SDMs, a formal performance
characterization process should be completed. This involves all advanced
PCB processing workflows, microfabrication/micropatterning, component
assembly on PCB/SDMs according to electromagnetic designs, PCB quality
control and lastly, their electrical characterization and wave control
performance assessment. Notably, the aforementioned processes are
non-standard in the emerging SDM field in general, and need to be
further specialized separately for SDM with and without impinging
wavefront capabilities.

\section{Conclusion}\label{sec:Conclusion}

Modern XR faces critical scalability, cost and complexity issues,
hindering its wide adoption. These shortcomings stem from the multitude
of involved devices that need to cooperate tightly across all layers
of the OSI stack. This present paper presented a new approach to XR
denoted as iCOPYWAVES, which dictates an end-to-end operation that
is bounded at the physical-layer, thereby yielding minimal system
latency and architectural simplicity. iCOPYWAVES is based on precise
RX imaging (denoted as XR-RF imaging), recent advances in 6G programmable
wireless propagation environments, and machine learning. We leverage
PWEs to selectively copy RF imaging wavefronts and wavevolumes from
one location in space to another, where a machine learning module,
accelerated by FPGAs, translates it to visual input for an XR headset.
The overarching ambition of iCOPYWAVES is to create a complete platform
for: i) creating RF wavefront representations of 3D objects, and ii)
manipulating these RF wavefronts with the ease of a \textquotedblleft copy-paste\textquotedblright{}
functionality--and no requirements for understanding the complex
underlying physics--thereby virtually transferring the original objects
to remote locations. The present study detailed the architecture and
end-to-end workflow of the iCOPYWAVES approach, and validated its
operation via simulations combining ray-tracing and generative adversarial
networks. Research challenges towards its implementation and new applications
enabled by iCOPYWAVES have been highlighted.

\section*{Acknowledgment}

This work was partially funded by the European Union's Horizon 2020
research and innovation programme-project COLLABS, GA EU871518, as
well as by the European Research Council (ERC) under grant agreement
No. 864228 (AdjustNet), 2020-2025 and the BMBF (Bundesministerium
f�r Bildung und Forschung) project, 6G-RIC: 6G Research and Innovation
Cluster, 2021-2025. 


\end{document}